\documentclass[fleqn,usenatbib]{mnras}

\usepackage{newtxtext,newtxmath}

\usepackage[T1]{fontenc}
\usepackage{ae,aecompl}

\usepackage{graphicx}	
\usepackage{amsmath}	
\usepackage{amssymb}	

\usepackage{natbib}
\usepackage{multicol}
\usepackage{lscape}

\usepackage[latin1]{inputenc}

\title[Properties of debris disc stars]{ Emerging trends  in metallicity  and lithium properties of debris disc stars
\thanks{Based on observations made with the 2.2 m telescope at the European Southern
Observatory (La Silla, Chile), under the agreement ESO-Observatorio Nacional/
MCTIC} }

\author[C. Chavero et al.]{
C. Chavero,$^{1}$\thanks{E-mail: carolina@oac.unc.edu.ar (C.CH)}
 R. de la Reza,$^{2}$
 L. Ghezzi,$^{2, 6}$
 F. Llorente de Andr\'es,$^{3}$
 C. B. Pereira,$^{2}$
\newauthor C. Giuppone,$^{4}$
and   G. Pinz\'on$^{5}$
\\
$^{1}$ Universidad Nacional de C\'ordoba, Observatorio Astron\'omico, Laprida 854, 5000 C\'ordoba, CONICET, Argentina\\
$^{2}$Observat\'orio Nacional, Rua General Jos\'e Cristino, 77, 20921-400, S\~ao Crist\'ov\~ao, Rio de Janeiro, RJ, Brazil\\
$^{3}$ Ateneo de  Almagro (Secc. Ciencia \& Tecnolog\'ia)- 13270 Almagro, Spain\\
$^{4}$ Universidad Nacional de C\'ordoba, Observatorio Astron\'omico, IATE, Laprida 854, 5000 C\'ordoba, Argentina\\
$^{5}$ Universidad Nacional de Colombia, Colombia\\
$^{6}$ Observat\'orio do Valongo, Universidade Federal do Rio de Janeiro, Ladeira do Pedro Ant\^onio 43, Rio de Janeiro, RJ 20080-090 
}

\date{Accepted 2019 May 21; Received 2019 May 21; in original form 2018 October 4}

 \pubyear{2019}

\begin{document}
\label{firstpage}
\pagerange{\pageref{firstpage}--\pageref{lastpage}}
\maketitle

\begin{abstract}
Dwarf stars with debris discs and planets appear to be excellent
laboratories to study the core accretion theory of planets formation.
These systems are however, insufficiently studied. In this paper we present 
the main metallicity and lithium abundance properties of these stars together
with stars with only debris discs and stars with only planets. Stars
without detected planets nor discs are also considered. The
 analysed sample is formed by main-sequence FGK field single
stars. Apart from the basic stellar parameters, we include the use of
dusty discs masses. The main results show for the first time that
the dust mass of debris disc stars with planets correlate with
metallicity.  We confirm that these disc dust masses are related to
their central stellar masses.

Separately, the masses of stars and those of planets
also correlate with metallicity. We conclude that two conditions are
necessary to form giant planets: to have a sufficient metallicity and
also a sufficient  protoplanetary mass of gas and dust. The
debris discs masses of stars without giant planets do not correlate
with metallicity, because they do not fulfil these two
conditions. Concerning lithium, by adopting a stellar model for
lithium depletion based on a strong interaction between the star
and a protoplanetary disc, we found that in agreement with the
model predictions, observations indicate that the main lithium
depletion occurs during this initial protoplanetary evolution
stage. We show that the ultimately lithium depletion is independent
of the presence or absence of planets and appears to be only age
dependent.
\end{abstract}

    


\section{Introduction}

If we consider the Solar System as an example of a debris disc, the planets coexist with  different kinds of debris represented by
asteroids, comets, minor bodies of the Kuiper Belt and zodiacal dust originated by disintegration of Jupiter family comets in the inner
Solar System  \citep{backmanParesce93,moromartin2002}. 
This example of dust--planets connection has been taken as a representative of what could be eventually found around stars.
This has been one the main leitmotifs of recent research, especially among FGK types of stars \citep{matthews2016}. However, debris discs detected
around stars are different from the Solar System case and present a large diversity in their properties. Two review papers devoted to stellar debris
discs appeared recently,  while \cite{hughes2018} discuss mainly their structure, composition and variability, \cite{wyatt2018} refers to aspects 
related to formation of low mass planets after the protoplanetary stage.

Debris discs  are mainly dusty structures that remain after the primordial protoplanetary (hereafter PP) disc has lost a large part of its original gas. A transition disc of short lifetime is established \citep{wyatt2015} before a typical dusty debris disc appears.
This debris disc is characterized by the presence  of an observed second generation of  dust grains produced continuously by collisional processes between 
larger bodies of kilometric sizes (planetesimals). These planetesimals were probably built when a stable and cool PP  disc    was formed 
\citep[see the reviews by][]{testi2014,drazkowska2018}. The evolution of PP and debris disc systems are then completely different. Whereas PP discs last from 3 Myr
to 10 Myr, the time necessary to be devoid of its main gas component, contrary to this, a debris disc can last practically the stellar lifetime. The PP  disc 
evolution is characterized by constructive processes where planetesimals, dust cores and gas giant planets are formed. In a debris system (DD), where some constructive
processes can exist forming low and very low mass planets, the evolution in a debris disc is mostly characterised by destructive collisional processes \citep{wyatt2005b,kenyon08}. The disc opacity, larger in the PP stage, is the main observational discriminator between PP and DD objects  \citep{hughes2018}.
Also, recent advances on the detection of gas in DD stars have been reported by these last authors, highlighting that the presence of gas appears specially
in AB type stars and much less in FGK type stars which are the objects of study of this work.

This paper is devoted to the study of two aspects concerning  stars with debris discs; their metallicity and their lithium (Li) evolution. The metallicity
properties of DD stars have been only partially tackled, leaving more questions than answers. As far as we know,  lithium abundance  properties in stars with
debris discs have never been studied in detail.

A study with a large important number of only FGK-type DD stars did not find trends with spectral type \citep{sierchio14}. As far as the FGK--types stars are
concerned, a large survey \citep{moromartin2015} with 200 single stars with ages larger than 1 Gyr has shown no clear planet -- debris disc relations.
However, this situation is now superseded by the recent detection of more  stars with debris disc and planet (DDP stars) which are now considered in this work. 
DDP stars contain all kind of planets up to a certain maximum upper limit.
Since the year 2012 have appeared some important works related to the possibility that debris discs, due to their dynamical stability during long periods of time, could host low mass terrestrial planets \citep{wyatt2012}. The \cite{raymond2011} simulations found positive correlations between debris discs and terrestrial planets. However, this correlation disappears in the presence of giant planets with eccentric orbits. This indicated positive correlation has been confirmed for low-mass planet by \cite{marshall2014}. As mentioned before, a recent review paper devoted to these aspects of low mass planets formation in debris disc stars, can be found in \cite{wyatt2018}. \cite{maldonado12} studied for the first time the most extended collection of these debris discs stars with planets.

As Li properties are concerned, we mean always the $^{7}$Li isotope. In this paper we study the mechanism of Li deptetion produced by the internal differential
stellar rotation as a function of a strong protoplanetary  (PP) disc magnetic interaction with the star during the short lifetime of this disc.
Several fundamental stellar parameters apart from the Li abundance (A(Li)) as the stellar mass ($M_{\star}$),  effective  temperatures ($T_{{\rm eff}}$), 
surface stellar rotation, measured here as the projected rotation velocity  ($\mathfrak{v}\sin i$), ages and metallicity ([Fe/H]) will be taken into account.
This ensemble of six parameters is homogeneously derived in our study for a large part of the stellar sample.

 This paper is organised as follows: in Section 2 we discuss on
our stellar sample, observations, stellar parameters determinations
and comparisons of our data with the literature. Section 3 is
devoted to the general metallicity properties of debris discs stars
with and without planets. In Section 4 are presented the main
lithium depletion properties in debris discs stars. Last, in Section 5
a discussion of the results and conclusions is given.

\section{Sample and observational data}

\subsection{The stellar sample}

 Our sample of  solar-like stars, with and without debris discs, is based on the presence or not of infrared (IR) excesses. 
 It was carefully compiled by checking mainly the works of;  \cite{trilling08}, \cite{bryden06}, \cite{hillenbrand08}, \cite{su05}, \cite{beichman06b}, \cite{gautier08},  \cite{hines06}, \cite{su06}, \cite{carpenter08}, \cite{smith06},
 and  \cite{matthews07}, which are mainly based in Spitzer data. The stellar sample of stars used in this work (full sample) is formed by a main collection of objects that have been observed and  reduced by us  (homogeneous sample)   using high-resolution echelle spectra (FEROS) to homogeneously determine some of the stellar properties. Also, in order to increase the number of  objects in the sample, we have added objects which parameters  were taken from the literature, compiling a total list of 140 stars.

Then, using the Extrasolar Planets Encyclopaedia\footnote{http://exoplanet.eu/  } \citep{schneider2011} we divide the samples considering the presence or not of planets.  Finally, we have obtained  four different groups of FGK type stars; 1) a control group formed by
38 stars with apparently no detected debris discs or planets. This group will be called as C in the whole paper. 2) a group of 41 stars containing only debris discs
(called DD) and not containing detected planets. 3) a group of  30  stars containing debris disc and planets of any mass, called DDP and finally, 4) a group formed 
by 31 stars with planets, called CP.  All observed and compiled stars in this study belong to spectral types, between F5 and K4 subtypes. In this way,
we avoid any hotter object than F4, that belongs to the Li-DIP phenomena  \citep{boesgaard86} indicating a different physical Li depletion process from
the one  considered in this work. Also, our selection of FGK subtypes characterises better what we can call ``solar'' low mass stars, where the theory of planet
formation is formulated.

In Table \ref{t:parameters0} we present the whole collection of the stars related to this work
with their stellar properties and references of the IR data. The stellar ages and masses
were derived using L. Girardi on-line code PARAM 1.3 \citep[http://stev.oapd.inaf.it/cgi-
bin/param, see also][]{dasilva2006}. This tool requires different input
parameters, such as: $\rm T_{{\rm eff}}$, [Fe/H], V magnitude and parallax, where the last two
parameters were taken from the {\sc Hipparcos} catalog \citep{ESA1997}. For $\rm T_{{\rm eff}}$, and [Fe/H] we used the
values  listed in Table \ref{t:parameters0}.

Concerning the debris discs studied here, a large part of them are obtained with data of the Spitzer Space Telescope by
means of MIPS photometry mainly  at 70 $\mu$m  \citep{chen2014}. The grains
responsibles for these  radiations are some unities of microns big.  Typical values of the ratios of luminosities (L$_{IR}$/L$_*$) are
of the order of 10$^{-6}$ to 10$^{-4}$. The sizes of the dust radii involve values from some unities of AU up to near 150 AU. In this study,
we also considered  the radiation observed at 850 $\mu$m for some stars. In this case, larger grains with sizes of more than
one mm are responsible for this radiation emitted at even larger distances, of one to ten times, the Kuiper Belt dust
radius.

\subsection{Observations and stellar parameters}

The high-resolution spectra of the stars  analysed in this work (74 stars) were obtained with the FEROS (Fiberfed Extended Range Optical Spectrograph)
echelle spectrograph \citep{Kaufer99} of the 2.2 m MPIA-ESO  telescope at La Silla (Chile). The FEROS spectral resolving power is R$=$48 000, corresponding to 2.2
pixel of $15~\mu$m, and the wavelength coverage ranges from 3500  to 9200~\AA. The nominal signal- to-noise ratio (S/N) was evaluated by measuring the rms-flux
fluctuation in selected continuum windows, and the typical values were S/N $\sim$ 250. The spectra were reduced automatically with the MIDAS pipeline reduction
package. 

Effective temperature ($\rm T_{{\rm eff}}$), surface gravity ($\log{g}$), microturbulence ($\xi_t$), metallicity ([Fe/H]), and  Li abundance  were derived by
means of the  standard approach of the local thermodynamic equilibrium (LTE) using a revised version (2002) of the code MOOG \citep{Sneden73} and a grid of
\cite{Kurucz93} ATLAS9 atmospheres, which include overshooting.

The atmospheric parameters were obtained from the equivalent width  of the  iron lines (Fe\,{\sc i} and Fe\,{\sc ii} )  by iterating until the correlation coefficients
between $\log\varepsilon$(Fe\,{\sc i}) and lower excitation potential ($\chi l$), and between  $\log\varepsilon$(Fe\,{\sc i}) and  reduced equivalent width 
($\log ({W_\lambda}/\lambda$) were zero, and the mean abundance given by Fe\,{\sc i} and Fe\,{\sc ii} lines were similar. The iron lines taken from \cite{lambert96}
and  \cite{santos04} were carefully chosen by verifying that each line was not too strong, and checked for possible blending. 
The equivalent widths were automatically measured with the ARES\footnote { http://www.astro.up.pt/$\sim$sousasag/ares}   code \citep{sousa07}.

 We  adopted new $\log{gf}$ values for the iron lines. These values were computed from an inverted solar analysis using solar  equivalent widths measured from a solar
 spectrum taken with FEROS and a Kurucz grid model for  the Sun (Kurucz 1993) having ($\rm T_{{\rm eff}}$, $\log{g}$, $\xi_{{\rm t}}$, $\log\varepsilon(\rm Fe)$)
 = (5777 K, 4.44 dex, 1.00 kms$^{-1}$, 7.47 dex). Table 1 of \cite{chavero2010} contains the  linelist used.

Final values of  the parameters  for all stars are presented in Table \ref{t:parameters0} and the columns are; the HD number, group  distinguishing the presence of disc or 
planet defined in the introduction, V magnitude, spectral type, stellar mass, age in Gyr, reference of the IR excess,  effective temperature in K, metalllicity [Fe/H]
\footnote{where  [Fe/H] = log $\epsilon$(Fe)$_{\star}$ - log $\epsilon$(Fe)$_{\odot}$}, stellar surface gravity as  log ${\it g}$  (g in cgs), the rotational velocity  $\mathfrak{v}\sin i$ in km $s^{-1}$, the Li abundance A(Li) = log N(Li)/N(H) $+$ 12 where N(Li) and N(H)  are the respective numbers of Li and H atoms, log of the mass
of the dusty disc which is defined in  Section~\ref{mdisc} and references of stellar parameters.

Uncertainties in the derived parameters  $\rm T_{{\rm eff}}$, $\log{g}$, $\xi_{{\rm t}}$, $\log\varepsilon(\rm Fe)$  were estimated as in \cite{gonzalezvanture98}.
The internal errors in the adopted effective temperature and microturbulence were determined from the uncertainty in the slope of  the Fe\,{\sc i} abundance versus 
excitation potential, and the Fe\,{\sc i} abundance versus reduced equivalent width relations, respectively. The uncertainty in $\log{g}$ was inferred by changing 
this parameter around the adopted solution until the Fe\,{\sc i} and  Fe\,{\sc ii} mean abundances differed by exactly one standard deviation of the mean value of the
Fe\,{\sc i} abundance. The typical values for the internal errors in this study are $\sim$ 70 K in $T_{{\rm eff}}$, 0.15 dex in $\log{g}$, 0.05 km s$^{-1}$ for $\xi_{{\rm t}}$, and 0.05 dex to 0.07 dex in [Fe/H].

The determination of Li abundances from synthetic spectra requires a line list for the spectral region around the Li {\sc i} feature
at 6707.8~\AA . The 2002 version of the code MOOG \citep{Sneden73} was used to
compute synthetic spectra in the Li region mentioned. The linelist and procedure follow the  methodology presented in detail in section 2 and 3 of \cite{ghezzi2010b}.
The formal uncertainties in the derived best-fit lithium abundances can be 
calculated by varying A(Li) around its best value and computing, for each lithium 
abundance tested, the quantity $\Delta\chi_{\rm r}^{2}$ = $\chi_{\rm r}^{2}$ - $\chi_{\rm r,min}^{2}$. The 
difference between A(Li) and A(Li)$_{best}$ that gives $\Delta\chi_{\rm r}^{2}$ = 1 is taken as the 1$\sigma$ 
uncertainty.

We estimated rotation velocities, $\mathfrak{v}\sin i$, by means of the spectral synthesis technique using determined atmosphere models. A few Fe\,{\sc I}
lines which fall in the same echelle order as the Li\,{\sc i} feature were investigated,  we chosen the Fe\,{\sc i} at 6703.567 \AA, it was identified as a 
clean line. Synthetic spectra were calculated using a macroturbulent velocity taken from \cite{valenti05} as first step. Then, a grid of synthetic spectra was
computed for combinations of  $\mathfrak{v}\sin i$ until to obtain the best fit. We corroborate that FEROS spectra are not necessarily sensitive to values
of  $\mathfrak{v}\sin i$ $<$ 2.5 km s$^{-1}$ for typical macroturbulent velocities, for this reason, for the most of slow rotating stars,  we could  only  obtain upper limit for this parameter. 

\begin{figure}
  \centering
  \includegraphics[width=7.0cm]{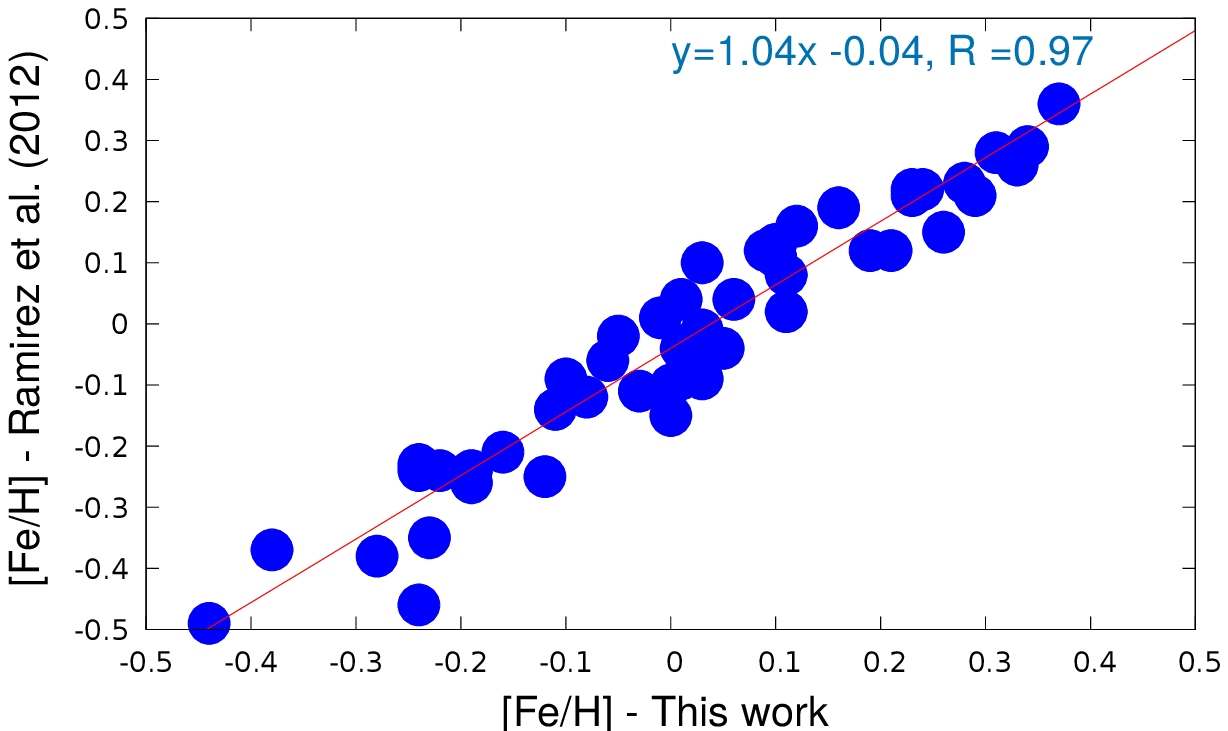}\par     
  \includegraphics[width=7.0cm]{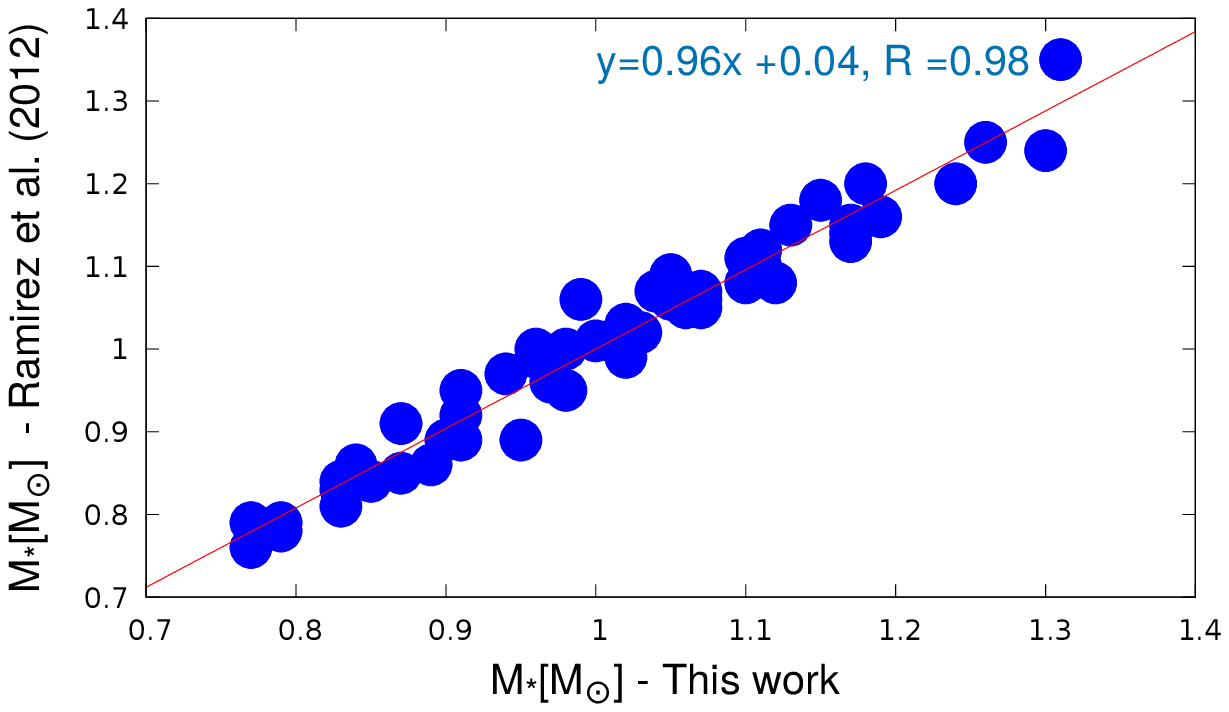}\par     
  \includegraphics[width=7.0cm]{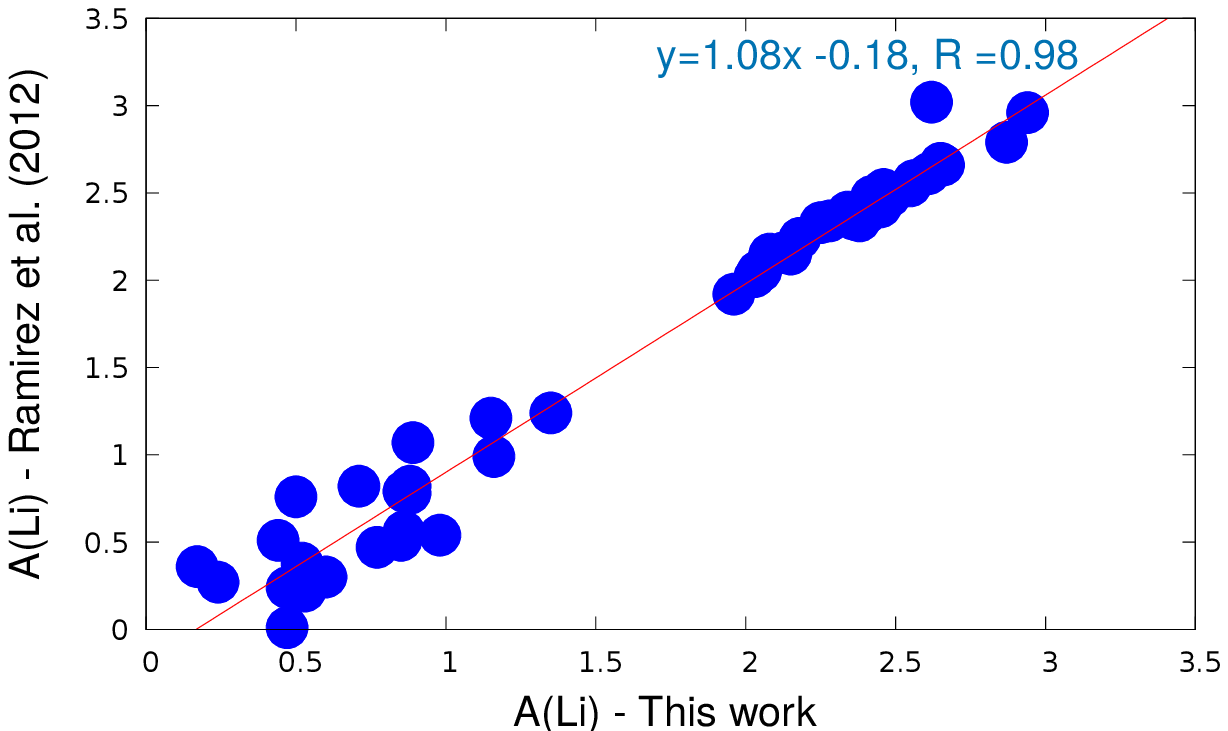}\par        
  \caption{  Comparison of our derived values of: a) [Fe/H], b) stellar mass (in solar mass units) and c)  A(Li), and those extracted from Ramirez et al. (2012). 
  Each panel contains the parameters of the linear fit and the value of the linear correlation coefficient R.}
  \label{ramirez}
   \end{figure}

 From the 140 stars of the whole sample studied in this paper, 74 of them were observed by us, and the stellar parameters, metallicities and A(Li) were calculated
 in this work as we describe above; the same parameters of 24 stars were taken from  \cite{ghezzi2010b},  which  use the same methodology for the lithium abundances determination as we mentioned before. We also considered 12 more stars from  \cite{ramirez2012}. For comparison purposes, Figure \ref{ramirez}  shows three  panels confronting our values and those obtained by \cite{ramirez2012} for stars in common: metallicities, stellar mass and  lithium abundances. The correlation coefficients
 show  that there is a very good agreement between the parameters obtained.  Taking into account the data taken from \cite{ghezzi2010b}, \cite{ramirez2012} 
 and derived in this work we attain the 77\% of the sample, for the rest of the stars we used data from literature as is specified in the last column of Table \ref{t:parameters0}.

\setcounter{table}{0}
\begin{table*}
\caption{Parameters of the stellar sample }
\label{t:parameters0}
\small
\begin{tabular}{lcccccccccccc}
\hline
\hline
 Star       &  Group   &      Sp.    &   $M_{\star}$             &  Age        & IR               & $T_{\rm eff}$    & [Fe/H] &  log ${\it g}$   &  $\mathfrak{v}\sin i$     & A(Li)  & log($M_{\rm d})$ & Ref\\                                                                                                                                                                                     
  HD        &          &       Type &     ($M_{\odot}$)        &  (Gyr)        &    Ref.          &   (K)            &  (dex) &     ($cgs$)       &(kms$^{-1}$)              &       &      ($M_{\rm moon}$) &                            \\
\hline                                                                                                                                                        
                                                                                                                                                          
 693     &        C    &  F5V          &  1.12 $\pm$ 0.03      &    4.7 $\pm$   0.7    &  (a)     &  6250     &   -0.24 $\pm$   0.04       &  4.20   &    6.50    &        2.45  $\pm$  0.11    &-- & TW                   \\
 3795    &        C    &  G3/5V        &  1.03 $\pm$ 0.02      &    9.3 $\pm$   0.4    &  (b)     &  5420     &   -0.54 $\pm$   0.06       &  3.94   &    $\leq$2 & $\leq$ 0.68                 &-- & TW                   \\
 3823    &        C    &  G1V          &  1.00 $\pm$ 0.02      &    7.8 $\pm$   0.8    &  (a)     &  5980     &   -0.23 $\pm$   0.06       &  4.10   &    3.5     &        2.36  $\pm$  0.11    &-- & TW                   \\
 4307    &        C    &  G2V          &  1.07 $\pm$ 0.02      &    7.1 $\pm$   0.5    &  (b)     &  5860     &   -0.14 $\pm$   0.07       &  4.00   &    $\leq$2 &        2.45  $\pm$  0.11    &-- & TW                   \\
 101501  &        C    &  G8V          &  0.93 $\pm$ 0.02      &    2.1 $\pm$   1.7    &  (a)     &  5528     &   -0.05 $\pm$   0.06       &  4.53   &      --    &        0.95 $\pm$ 0.02     &-- & R12             \\
 10476   &        C    &  K1V          &  0.83 $\pm$ 0.02      &    5.0 $\pm$   4.0    &  (c)     &  5180     &   -0.05 $\pm$   0.06       &  4.38   &  $\leq$ 2  & $\leq$ 0.24                 &-- & TW           \\
 11131   &        C    &  G0           &  1.02 $\pm$ 0.02      &    1.5 $\pm$   1.4    &  (a)     &  5860     &   0.01  $\pm$   0.07       &  4.50   &    2.5     &        2.55  $\pm$  0.11    &-- & TW           \\
 14412   &        C    &  G8V          &  0.77 $\pm$ 0.02      &    5.0 $\pm$   4.0    &  (a)     &  5400     &   -0.44 $\pm$   0.09       &  4.56   &    3.5     & $\leq$ 0.88    0.34         &-- & TW           \\
 14802   &        C    &  G2V          &  1.17 $\pm$ 0.04      &    5.4 $\pm$   0.8    &  (a)     &  5920     &   0.02  $\pm$   0.06       &  4.14   & $\leq$ 2   &        2.38  $\pm$  0.11    &-- & TW           \\
 20630   &        C    &  G5V          &  0.98 $\pm$ 0.03      &    2.7 $\pm$   2.4    &  (b)     &  5680     &   0.06  $\pm$   0.04       &  4.40   &    4.5     &        2.04  $\pm$  0.11    &-- & TW           \\
 20766   &        C    &  G2V          &  0.91 $\pm$ 0.03      &    4.1 $\pm$   3.3    &  (b)     &  5715     &   -0.19 $\pm$   0.06       &  4.50   &$\leq$ 2    & $\leq$ 0.77                 &-- & TW            \\
 21197   &        C    &  K4V          &  0.76 $\pm$ 0.03      &    1.5                &  (a)     &  4815     &   0.18  $\pm$   0.06       &  4.60   &     0.53   &        0.11               &-- & S06.L17             \\
 26923   &        C    &  G0IV         &  1.06 $\pm$ 0.02      &    1.3 $\pm$   1.2    &  (a)     &  5980     &   0.04  $\pm$   0.07       &  4.42   &    3.5     &        2.75  $\pm$  0.10    &-- & TW            \\
 30652   &        C    &  F6V          &  1.28 $\pm$ 0.03      &    1.8 $\pm$   1.0    &  (a)     &  6477     &   0.00  $\pm$   0.06       &  4.29   &      --    &        2.12     $\pm$ 0.04     &-- & R12            \\
 34721   &        C    &  G0V          &  1.09 $\pm$ 0.03      &    5.3 $\pm$   1.1    &  (a)     &  6020     &   -0.02 $\pm$   0.06       &  4.30   &$\leq$ 2.5  &        2.34  $\pm$  0.11    &-- & TW            \\
 35296   &        C    &  F8V          &  1.11 $\pm$ 0.03      &    3.2 $\pm$   1.6    &  (a)     &  6080     &   0.00  $\pm$   0.07       &  4.20   &    16      &        2.94  $\pm$  0.11    &-- & TW            \\
 43162   &        C    &  G5V          &  0.94 $\pm$ 0.03      &    2.8 $\pm$   2.7    &  (d)     &  5600     &   -0.01 $\pm$   0.04       &  4.40   &    5       &        2.25  $\pm$  0.11    &-- & TW              \\
 43834   &        C    &  G5V          &  0.97 $\pm$ 0.03      &    3.9 $\pm$   3.1    &  (a)     &  5580     &   0.13  $\pm$   0.06       &  4.60   & $\leq$ 2   & $\leq$ 0.71                 &-- & TW             \\
 63333   &        C    &  F5           &  0.99 $\pm$ 0.03      &    6.5 $\pm$   1.0    &  (a)     &  6178     &   -0.38 $\pm$   0.06       &  4.25   &    3.0     &        2.46  $\pm$  0.11    &-- & TW             \\
 71640   &        C    &  F5           &  1.04 $\pm$ 0.03      &    5.1 $\pm$   1.7    &  (a)     &  6060     &   -0.13 $\pm$   0.07       &  4.26   &    4.5     &        2.44  $\pm$  0.11    &-- & TW             \\
 76932   &        C    &  G2V          &  0.85 $\pm$ 0.03      &   11.8 $\pm$  11.0    &  (d)     &  5959     &   -0.83 $\pm$   0.04       &  4.16   &       --   &        2.12    $\pm$ 0.02     &-- & R12             \\
 122862  &        C    &  G1V          &  1.05 $\pm$ 0.02      &    7.3 $\pm$   0.6    &  (n)     &  5900     &   -0.11 $\pm$   0.06       &  4.00   &    3.5     &        2.40  $\pm$  0.12    &-- & TW             \\
 133295  &        C    &  G0/1V        &  1.09 $\pm$ 0.02      &    1.2 $\pm$   1.0    &  (b)     &  6060     &   0.03  $\pm$   0.06       &  4.50   &    10      &        2.84  $\pm$  0.11    &-- & TW             \\
 142267  &        C    &  G1V          &  0.87 $\pm$ 0.02      &    8.9 $\pm$   2.3    &  (b)     &  5740     &   -0.3  $\pm$   0.07       &  4.52   & $\leq$ 2   & $\leq$ 0.92                 &-- & TW            \\
 142860  &        C    &  F6IV         &  1.23 $\pm$ 0.03      &    4.5 $\pm$   3.1    &  (a)     &  6313     &   -0.19 $\pm$   0.06       &  4.20   &     --     &        2.12     $\pm$ 0.03     & -- & R12             \\
 149661  &        C    &  K2V          &  0.85 $\pm$ 0.02      &    3.5 $\pm$   3.3    &  (d)     &  5280     &   -0.01 $\pm$   0.06       &  4.50   & $\leq$ 2   & $\leq$ 0.47                &-- & TW             \\
 152391  &        C    &  G8V          &  0.91 $\pm$ 0.02      &    2.4 $\pm$   2.4    &  (h)     &  5500     &   0.01  $\pm$   0.07       &  4.56   &    2.5     &        1.35  $\pm$  0.21    &-- & TW             \\
 165499  &        C    &  G0V          &  1.07 $\pm$ 0.03      &    4.8 $\pm$   1.6    &  (h)     &  5980     &   0.00  $\pm$   0.06       &  4.30   &    2.5     &        2.05  $\pm$  0.14    &-- & TW             \\
 172051  &        C    &  G5V          &  0.87 $\pm$ 0.03      &    4.6 $\pm$   3.6    &  (h)     &  5620     &   -0.22 $\pm$   0.06       &  4.46   & $\leq$ 2   &        1.15  $\pm$  0.22    &-- & TW             \\
 177565  &        C    &  G8V          &  0.96 $\pm$ 0.03      &    5.6 $\pm$   3.5    &  (a)     &  5580     &   0.10  $\pm$   0.07       &  4.34   & $\leq$ 2   & $\leq$ 0.44                 &-- & TW             \\
 181321  &        C    &  G1/2V        &  1.03 $\pm$ 0.02      &    1.4 $\pm$   1.3    &  (b)     &  5860     &   0.05  $\pm$   0.06       &  4.56   &    12      &        2.95  $\pm$  0.11    &-- & TW             \\
 187691  &        C    &  F8V          &  1.26 $\pm$ 0.02      &    2.6 $\pm$   0.5    &  (a)     &  6160     &   0.19  $\pm$   0.05       &  4.32   &    3.5     &        2.65  $\pm$  0.11    & --& TW             \\
 190248  &        C    &  G5IV         &  1.05 $\pm$ 0.02      &    5.9 $\pm$   2.0    &  (a)     &  5600     &   0.37  $\pm$   0.06       &  4.25   &$\leq$ 2    & $\leq$ 0.98                 & --& TW             \\
 196761  &        C    &  G8V          &  0.83 $\pm$ 0.02      &    5.7 $\pm$   4.0    &  (i)     &  5440     &   -0.24 $\pm$   0.06       &  4.44   &$\leq$ 2    & $\leq$ 0.52                 & --& TW             \\
 200968  &        C    &  K1V          &  0.84 $\pm$ 0.02      &    5.1 $\pm$   4.1    &  (a)     &  5140     &   0.00  $\pm$   0.06       &  4.32   &$\leq$ 2    & $\leq$ 0.36                 & --& TW             \\
 203608  &        C    &  F6V          &  0.91 $\pm$ 0.03      &    6.7 $\pm$   1.9    &  (a)     &  6160     &   -0.53 $\pm$   0.07       &  4.42   &    3.5     &        2.48  $\pm$  0.11    & --& TW             \\
 210918  &        C    &  G5V          &  0.98 $\pm$ 0.03      &    7.8 $\pm$   2.0    &  (b)     &  5780     &   -0.03 $\pm$   0.06       &  4.36   & $\leq$ 2   & $\leq$ 0.53                 & --& TW             \\
 212330  &        C    &  F9V          &  1.24 $\pm$ 0.04      &    4.4 $\pm$   0.6    &  (b)     &  5700     &   0.05  $\pm$   0.06       &  4.3    & $\leq$ 2.5 &        1.60  $\pm$  0.18    & --&TW            \\
    \hline  
 1237      &        CP      &  G6V       &  0.94 $\pm$ 0.03       &    2.9 $\pm$ 2.7   &    (b)   &    5500     &   0.09 $\pm$   0.03  &4.40   & 4.50      &       2.08   $\pm$ 0.11    & -- &    TW    \\                     
 3651      &        CP      &  K0V       &  0.89 $\pm$ 0.03       &    5.0 $\pm$ 4.0   &    (b)   &    5240     &   0.12 $\pm$   0.04  &4.32   & 2.50      &$\leq$ 0.60                 & -- &    TW    \\                      
 4308      &        CP      &  G3V       &  0.87 $\pm$ 0.06       &    7.1 $\pm$ 3.0   &    (o)   &    5644     &  -0.34 $\pm$   0.06  &4.38   & --        &       1.04   $\pm$ 0.26    & -- &    G10    \\                      
 10697     &        CP      &  G3Va      &  1.14 $\pm$ 0.03       &    6.7 $\pm$ 0.8   &    (a)   &    5650     &   0.18 $\pm$   0.03  &4.10   & 2.00      &       1.97   $\pm$ 0.11    & -- &    TW    \\                      
 13445     &        CP      &  K0V       &  0.79 $\pm$ 0.02       &    6.5 $\pm$ 4.1   &    (a)   &    5140     &  -0.24 $\pm$   0.06  &4.46   & 2.00      &$\leq$ 0.17                 & -- &    TW    \\                      
 17051     &        CP      &  G0V       &  1.20 $\pm$ 0.01       &    0.4 $\pm$ 0.2   &    (b)   &    6239     &   0.12 $\pm$   0.07  &4.55   & 5.50      &       2.52   $\pm$ 0.11    & -- &    G10    \\  
 23079     &        CP      &  G0V       &  1.03 $\pm$ 0.04       &    3.9 $\pm$ 2.2   &    (a)   &    5990     &  -0.08 $\pm$   0.06  &4.50   & 3.00      &       2.17   $\pm$ 0.11    & -- &    TW    \\                      
 28185     &        CP      &  G5        &  1.02 $\pm$ 0.03       &    4.6 $\pm$ 3.2   &    (a)   &    5620     &   0.23 $\pm$   0.02  &4.30   & 2.50      &$\leq$ 0.50                 & -- &    TW    \\ 
 33564     &        CP      &  F6V       &  1.25 $\pm$ 0.04       &    3.0 $\pm$ 0.3   &    (o)   &    6250     &   0.12 $\pm$   0.05  &4.00   & --        &       2.30   $\pm$ 0.12    & -- &        GO10    \\ 
 72659     &        CP      &  G0        &  1.03 $\pm$ 0.02       &    8.3 $\pm$ 0.4   &    (b)   &    5851     &  -0.11 $\pm$   0.04  &4.01   & 0.50      &       2.16   $\pm$ 0.11    &  --&    G10    \\
 75732     &        CP      &  G8V       &  0.93 $\pm$ 0.09       &    5.1 $\pm$ 2.7   &    (o)   &    5279     &   0.33 $\pm$   0.02  &4.37   & --        &       0.35   $\pm$ 0.08    &  --&    R12    \\                     
 95128     &        CP      &  G0V       &  1.08 $\pm$ 0.04       &    7.4 $\pm$ 1.9   &    (o)   &    5954     &   0.06 $\pm$   0.03  &4.44   & --        &       1.74   $\pm$ 0.04    &  --&    R12    \\                     
 102365    &        CP      &  G3/5V     &  0.86 $\pm$ 0.03       &    8.5 $\pm$ 2.8   &    (g)   &    5665     &  -0.35 $\pm$   0.03  &4.42   & 0.20      &$\leq$ 0.52                 &  --&    G10    \\                     
 114729    &        CP      &  G0V       &  0.92 $\pm$ 0.01       &    10.8$\pm$ 0.5   &    (b)   &    5779     &  -0.33 $\pm$   0.06  &4.01   & 4.20      &       1.88   $\pm$ 0.11    &  --&    G10    \\                     
 115383    &        CP      &  G0V       &  1.15 $\pm$ 0.03       &    3.0 $\pm$ 1.0   &    (o)   &    6133     &   0.24 $\pm$   0.07  &4.63   & --        &       2.84   $\pm$ 0.10    &  --&    G10    \\                     
 134987    &        CP      &  G5V       &  1.11 $\pm$ 0.03       &    3.7 $\pm$ 1.9   &    (a)   &    5780     &   0.31 $\pm$   0.08  &4.24   & 2.00      &$\leq$ 0.89                 &  --&    TW    \\                     
 136352    &        CP      &  G2V       &  0.87 $\pm$ 0.02       &    9.6 $\pm$ 1.8   &    (b)   &    5700     &  -0.28 $\pm$   0.07  &4.34   & 2.00      &$\leq$ 0.86                 & -- &    TW    \\                     
 145675    &        CP      &  K0V       &  0.97 $\pm$ 0.01       &    4.6 $\pm$ 1.5   &    (o)   &    5311     &   0.43 $\pm$   0.03  &4.42   & --        &       1.19   $\pm$ 0.15    & -- &    L06     \\
 147513    &        CP      &  G3/5 V    &  1.06 $\pm$ 0.02       &    1.0 $\pm$ 0.9   &    (b)   &    5920     &   0.11 $\pm$   0.04  &4.42   & 2.00      &       2.00   $\pm$ 0.14    & -- &    TW     \\                 
 154088    &        CP      &  K1V       &  0.94 $\pm$ 0.07       &    6.4 $\pm$ 2.2   &    (b)   &    5374      &   0.28 $\pm$   0.06  &4.37  & --        &        --                  & -- &    S08     \\                 
 154345    &        CP      &  G8V       &  0.90 $\pm$ 0.12       &    4.1 $\pm$ 1.2   &    (o)   &    5468     &  -0.10 $\pm$   0.07  &4.54   & --        &$\leq$ -0.16                & -- &    GO10.R12    \\      
 160691   &        CP       &  G5V       &  1.13 $\pm$ 0.03       &    5.0 $\pm$ 1.2   &    (a)   &    5760    &   0.33 $\pm$   0.03  &4.24    & 2.50      &       1.16   $\pm$ 1.08    & -- &    TW    \\ 
\hline                                                                                                                                           
\end{tabular}                                                                                                                                             
\end{table*}

\setcounter{table}{0}
\begin{table*}
\caption{Parameters of the stellar sample }
\label{t:parameters0}
\small
\begin{tabular}{lcccccccccccc}
\hline
\hline                                                                                                                                             
  Star      &  Group   &       Sp. &   $M_{\star}$             &  Age    & IR                      & T$_{\rm eff}$ & [Fe/H]  &  log ${\it g}$   &  $\mathfrak{v}\sin i$    & A(Li)  & Log($M_{\rm d})$  & Ref\\                                     
   HD         &          &      Type         &     ($M_{\odot}$)     &  (Gyr)                &    Ref.                &   (K)         &  (dex)  &     ($cgs$)             &(kms$^{-1}$)     &       &    ($M_{\rm moon}$)  &     \\
\hline                                                                                                                                              
  189567             &        CP      &  G2V       &  0.90 $\pm$ 0.06       &    13.0$\pm$ 1.0   &    (o)   &    5726     &   0.24 $\pm$  0.04   &4.41   & --     & $\leq$ 0.86      & --  &   TW      \\ 
  189733             &        CP      &  G5        &  0.82 $\pm$ 0.02       &    2.2 $\pm$ 2.4   &    (b)   &    5201     &  -0.02 $\pm$  0.01   &4.64   & 0.80   & $\leq$ 0.08    & --  &   G10     \\ 
  192310             &        CP      &  K3V       &  0.80 $\pm$ 0.07       &    8.1 $\pm$ 3.2   &    (b)   &    5099     &  -0.03 $\pm$  0.06   &4.43   & --     & $\leq$ 0.24    & --  &   G10     \\ 
  195019             &        CP      &  G3V       &  1.03 $\pm$ 0.01       &    8.8 $\pm$ 0.3   &    (b)   &    5741     &  -0.01 $\pm$  0.04   &4.06   & 4.20   &        1.45   $\pm$     0.16  & --  &   G10      \\ 
  196050             &        CP      &  G3V       &  1.18 $\pm$ 0.02       &    2.5 $\pm$ 1.3   &    (a)   &    5920     &   0.34 $\pm$  0.06   &4.32   & 3.00   &        2.15   $\pm$      0.12  &  -- &   TW       \\
  196885             &        CP      &  F8V       &  1.26 $\pm$ 0.02       &    1.9 $\pm$ 0.6   &    (b)   &    6218     &   0.15 $\pm$  0.07   &4.18   & 7.90   &        2.64   $\pm$     0.11  &  -- &   G10      \\ 
  213240             &        CP      &  G4IV      &  1.19 $\pm$ 0.04       &    4.7 $\pm$ 0.9   &    (a)   &    5930     &   0.16 $\pm$  0.05   &4.20   & 3.50   &        2.41   $\pm$      0.11  &  -- &   TW       \\ 
  216437             &        CP      &  G1V       &  1.17 $\pm$ 0.04       &    5.0 $\pm$ 1.0   &    (a)   &    5840     &   0.24 $\pm$  0.06   &4.16   & 2.00   &        1.93   $\pm$      0.11  &  -- &   TW       \\ 
  217014             &        CP      &  G5V       &  1.05 $\pm$ 0.03       &    5.8 $\pm$ 1.5   &     (b)  &    5739     &   0.15 $\pm$  0.07   &4.17   & 2.80   &        1.24   $\pm$     0.22  &  -- &   G10      \\ 
 \hline                                                                                                                                                                      
  166                &        DD      &  K0V       &  0.95 $\pm$ 0.02      &    1.9  $\pm$ 1.8     & (b)  &   5465      &   0.14  $\pm$   0.04  & 4.53  &    4.1         &        2.38 $\pm$ 0.04 &   -2.48$^\dagger$   & R12     \\ 
  377                &        DD      &  G2V       &  1.02 $\pm$ 0.02          &    0.2  $\pm$ 0.2     & (k)  &   5873      &   0.13  $\pm$   0.07  & 4.28  &    14.6        &        2.90 $\pm$ 0.05 &   -0.82$^\dagger$   & V05.A18  \\ 
  1581               &        DD      &  F9V       &  0.96 $\pm$ 0.03      &    6.8  $\pm$ 2.4     & (a)  &   5900      &  -0.15  $\pm$   0.06  & 4.26  &    3.0         &        2.18 $\pm$ 0.11 &   -3.05$^*$   & TW      \\ 
  1835               &        DD      &  G3V       &  1.07 $\pm$ 0.02      &    1.2  $\pm$ 1.0     & (f)  &   5810      &   0.23  $\pm$   0.06  & 4.40  &    7.0         &        2.56 $\pm$ 0.11 &   -3.23$^\dagger$   & TW      \\ 
  7570               &        DD      &  F8V       &  1.15 $\pm$ 0.03      &    4.5  $\pm$ 1.0     & (a)  &   6190      &   0.18  $\pm$   0.07  & 4.45  &    4.3         &        2.87 $\pm$ 0.11 &   -3.72$^\dagger$   & TW      \\ 
  10008              &        DD      &  G5        &  0.86 $\pm$ 0.03      &    2.6  $\pm$ 2.6     & (c)  &   5360      &   -0.01 $\pm$   0.07  & 4.54  &    2.1         &        2.25 $\pm$ 0.11 &   -1.77$^\dagger$   & TW      \\ 
  17925              &        DD      &  K1V       &  0.84 $\pm$ 0.03      &    3.6  $\pm$ 3.4     &  (a) &   5180      &   0.03  $\pm$   0.06  & 4.44  &    6.3         &        2.62 $\pm$ 0.10 &   --  & TW      \\ 
  22484              &        DD      &  F9V       &  1.17 $\pm$ 0.04      &    5.1  $\pm$ 0.9     &  (a) &   6050      &   0.04  $\pm$   0.07  & 4.20  &    4.2         &        2.40 $\pm$ 0.11 &   -3.40$^\dagger$   & TW      \\ 
  30495              &        DD      &  G3V       &  1.01 $\pm$ 0.03      &    2.3  $\pm$ 2.0     & (a)  &   5795      &   0.03  $\pm$   0.06  & 4.41  &    3.0         &        2.34 $\pm$ 0.11 &   -2.15$^\dagger$   & TW      \\ 
  33262              &        DD      &  F7V       &  1.08 $\pm$ 0.03      &    1.1  $\pm$ 1.2     &  (a) &   6147      &  -0.17  $\pm$   0.06  & 4.44  &    15.4        &        3.09 $\pm$ 0.15 &   --   & R12     \\ 
  33636              &        DD      &  G0        &  1.02 $\pm$ 0.03      &    1.3                & (f)  &   6000      &  -0.10  $\pm$   0.04  & 4.62  &    1.3         &        2.40 $\pm$ 0.11 &   -1.52$^\dagger$   & L06     \\ 
  61005              &        DD      &  G3/5V     &  0.90 $\pm$ 0.03      &    2.7  $\pm$ 2.7     & (k)  &   5480      &   0.00  $\pm$   0.02  & 4.35  &    9.1         &        2.79 $\pm$ 0.11 &    0.15$^\dagger$   & TW       \\ 
  72905              &        DD      &  G15Vb     &  1.01 $\pm$ 0.02      &    1.7  $\pm$ 1.5     & (f)  &   5876      &  -0.07  $\pm$   0.04  & 4.49  &    9.6         &        2.94 $\pm$ 0.09 &   -2.89$^\dagger$   & R12      \\ 
  73350              &        DD      &  G0        &  1.06 $\pm$ 0.03      &    1.1                & (c)  &   5815      &   0.14  $\pm$   0.08  & 4.44  &    3.55        &        2.35 $\pm$ 0.10 &      & T10.R12   \\ 
  76151              &        DD      &  G3V       &  1.04 $\pm$ 0.03      &    2.6                & (g)  &   5770      &   0.11  $\pm$   0.04  & 4.40  &    3.9         &        1.77 $\pm$ 0.11 &   -3.40$^\dagger$   & G10      \\ 
  85301              &        DD      &  G5        &  1.02 $\pm$ 0.02      &    0.7  $\pm$ 0.5     & (f)  &   5782      &   0.13  $\pm$   0.05  & 4.56  &    6.2         &        2.02 $\pm$ 0.08 &   -1.62$^\dagger$   & G15     \\ 
  104860             &        DD      &  F8        &  1.04 $\pm$ 0.03       &    3.6  $\pm$ 2.4     & (k)  &   5956      &  -0.18  $\pm$   0.08  & 4.43  &    15          &          --            &    0.18$^\dagger$   & C11      \\ 
  107146             &        DD      &  G2V       &  1.02 $\pm$ 0.03       &    3.2  $\pm$ 2.2     & (k)  &   5882      &  -0.16  $\pm$   0.06  & 4.46  &    4.9         &        3.02 $\pm$ 0.04 &    0.08$^\dagger$   & V05.A18   \\ 
  110897             &        DD      &  G0V       &  0.96 $\pm$ 0.03       &    5.5  $\pm$  3.8    & (a)  &   5959      &  -0.59  $\pm$   0.05  & 4.43  &    0.1         &        1.94 $\pm$ 0.11 &   -1.96$^\dagger$   & L06     \\ 
  111347             &        DD      &  F7V       &  1.24 $\pm$ 0.03       &    1.1                & (f)  &   6471      &  -0.10   $\pm$   0.05 & 4.35  &    49.4        &          --            &   -3.34$^\dagger$   & C11      \\    
  118972             &        DD      &  K1V       &  0.83 $\pm$ 0.02       &    3.4  $\pm$ 3.3     & (a)  &   5240      &  -0.06  $\pm$   0.08  & 4.36  &    5.2         & $\leq$ 0.85            &   -3.32$^\dagger$   & TW       \\ 
  119124             &        DD      &  F7V       &  1.09 $\pm$ 0.03       &    4.1  $\pm$  3.0    & (f)  &   6215      &  -0.31  $\pm$   0.07  & 4.67  &    10.8        &    --                  &   -1.74$^\dagger$   & M15      \\ 
  129590             &        DD      &  G3V       &  1.30 $\pm$ 0.03       &    0.015              & (f)  &   5945      &  -0.09  $\pm$   0.04  & --    &    32          &    --                  &   -1.24$^\dagger$   & G16.M17   \\ 
  135599             &        DD      &  K0        &  0.82 $\pm$ 0.02       &    3.8  $\pm$ 3.5     & (c)  &   5220      &  -0.09  $\pm$   0.04  & 4.34  &    3.7         & $\leq$ 0.47            &   -1.80$^\dagger$   & TW        \\ 
  142446             &        DD      &  F3V       &  1.40 $\pm$ 0.03       &    0.016              & (f)  &   6710      &   0.12  $\pm$   0.08  & --    &    66          &    --                  &   -0.31$^\dagger$   & G16.C16   \\ 
  151044             &        DD      &  F8V       &  1.11 $\pm$ 0.03       &    3.8  $\pm$ 1.2     & (f)  &   6110      &  -0.09  $\pm$   0.08  & 4.31  &    5.2         &        2.72 $\pm$ 0.03 &   -1.48$^\dagger$ &V05.A18.B91   \\ 
  158633             &        DD      &  K0V       &  0.76 $\pm$ 0.03       &    6.3  $\pm$         & (f)  &   5330      &  -0.45  $\pm$   0.07  & 4.55  &    0.1         & $\leq$ 0.11            &   -2.30$^\dagger$   & R12      \\ 
  170773             &        DD      &  F5V       &  1.35 $\pm$ 0.03       &    0.9  $\pm$ 0.6     & (f)  &   6694      &  -0.04  $\pm$   0.07  & 4.27  &    67.2        &    --                  &    0.51$^\dagger$   & C11      \\ 
  181327             &        DD      &  F5/6V     &  1.24 $\pm$ 0.03       &     1.8  $\pm$ 0.6    & (f)  &   6541      &  -0.14 $\pm$    0.08  & 4.48  &    21          &        3.36            &    -0.11$^\dagger$   & S09.C11   \\ 
  183216             &        DD      &  G2V       &  1.17 $\pm$ 0.03       &    0.7  $\pm$0.6      & (n)  &   6080      &   0.25 $\pm$   0.07   & 4.58  &  5.4           &        2.72 $\pm$  0.11 &   -2.56$^\dagger$   & TW       \\ 
  187897             &        DD      &  G5        &  1.12 $\pm$ 0.03      &    1.3  $\pm$ 1.1     & (k)  &   5960      &   0.16  $\pm$   0.06  & 4.40  &  3.9           &        2.50 $\pm$  0.11 &   -1.28$^\dagger$   & TW       \\ 
  191089             &        DD      &  F5V       &  1.27 $\pm$ 0.03       &    0.024  $\pm$ 0.03         & (k)  &   6510      &  -0.09  $\pm$   0.05  & 4.28  &  37.7          &        3.21             &   0.75 $^\dagger$   & S09.C11.B15   \\ 
  193017             &        DD      &  F8        &  1.13 $\pm$ 0.03      &    0.8  $\pm$ 0.7     & (n)  &   6140      &   0.08  $\pm$   0.07  & 4.44  &  5.4           &        2.89 $\pm$ 0.11 &   -1.64$^*$   & TW       \\ 
  201219             &        DD      &  G5        &  0.99 $\pm$ 0.03      &    2.0  $\pm$ 2.0     & (k)  &   5620      &   0.15  $\pm$   0.07  & 4.40  &  2.9           &        1.16 $\pm$ 0.29 &   -0.89$^\dagger$   & TW       \\ 
  202628             &        DD      &  G5V       &  1.01 $\pm$ 0.03      &    2.6  $\pm$ 2.3     & (i)  &   5780      &   0.03  $\pm$   0.07  & 4.46  &  2.64          &        2.16 $\pm$ 0.11 &     & TW       \\ 
  202917             &        DD      &  G5V       &  0.95 $\pm$ 0.03      &    1.9  $\pm$ 1.9     & (k)  &   5680      &   0.00  $\pm$   0.08  & 4.50  &  14.7          &        3.34 $\pm$ 0.11 &   -1.89$^\dagger$   & TW       \\ 
   205536            &        DD      &  G8V       &  0.95 $\pm$ 0.03      &    5.4  $\pm$ 3.9     & (l)  &   5460      &  -0.04 $\pm$   0.08  & 4.38  &  1.8           & $\leq$ 0.42         &   -3.28$^*$   & TW       \\ 
   205905            &        DD      &  G2V       &  1.08 $\pm$ 0.02      &    1.1  $\pm$ 0.9     & (n)  &   5940      &   0.13 $\pm$   0.06  & 4.42  &  2.3           &        2.39 $\pm$ 0.11 &   -1.99$^\dagger$   & TW       \\ 
   207129            &        DD      &  G2V       &  1.04 $\pm$ 0.04      &    2.9  $\pm$ 2.2     & (a)  &   5900      &   0.01 $\pm$   0.08  & 4.40  &  2.71          &        2.28 $\pm$ 0.11 &    0.34$^\dagger$   & TW       \\ 
   209253            &        DD      &  F6/7V     &  1.14 $\pm$ 0.03      &     1.1                & (f)  &   6280      &  -0.16$\pm$   0.04  & 4.42  &  16.1          &        2.84  $\pm$ 0.03 &    -1.43$^\dagger$   & V05.A18   \\ 
   219482            &        DD      &  F7V       &  1.12 $\pm$ 0.03      &    4.7                & (f)  &   6249      &  -0.21 $\pm$   0.07  & 4.36  &  9.0           &         --    &   -2.63$^\dagger$   & E12       \\ 
  \hline 
  1461               &        DDP     &  G0V       &  1.05 $\pm$ 0.03      & 4.3  $\pm$  2.5     &  (f)  &   5740      &   0.18  $\pm$ 0.04  &  4.39  & 1.8        &        0.49 $\pm$ 0.08  &    -1.58$^\dagger$ &   R12   \\ 
  10647              &        DDP     &  F8V       &  1.10 $\pm$ 0.03      & 1.6  $\pm$  1.3     &  (g)  &   6155      &   -0.05 $\pm$ 0.04  &  4.44  & 5.3        &        2.74 $\pm$ 0.10  &     0.04$^\dagger$ &   G10    \\ 
  10700              &        DDP     &  G8V       &  0.77 $\pm$ 0.02      & 8.2  $\pm$  3.2     &  (j)  &   5344      &   -0.52 $\pm$ 0.07  &  4.50  &  --        & $\leq$ 0.42             &    -4.64$^*$ &   TW    \\
  20794              &        DDP     &  G8V       &  0.81 $\pm$ 0.03      & 11.4 $\pm$  0.1     & (b)   &   5401      &   -0.41 $\pm$ 0.03  &  4.40  &  --        & $\leq$ 0.58             &    -5.16$^*$ &   G10    \\
  22049              &        DDP     &  K2V       &  0.79 $\pm$ 0.02      & 3.8  $\pm$  3.6     &  (e)  &   5100      &   -0.15 $\pm$ 0.06  &  4.44  & 2.45       & $\leq$ 0.47             &    -3.66$^*$ &   TW    \\
  38858              &        DDP     &  G4V       &  0.89 $\pm$ 0.02      & 7.3  $\pm$  1.4     &  (b)  &   5733      &   -0.22 $\pm$ 0.04  &  4.51  & 0.1        &        1.48 $\pm$ 0.02  &    -1.35$^\dagger$ &   T10.I09    \\
  39091              &        DDP     &  G0V       &  1.10 $\pm$ 0.03      & 3.0  $\pm$  1.8     &  (e)  &   5950      &   0.11  $\pm$ 0.06  &  4.30  & 2.96       &        2.25 $\pm$ 0.11  &    -2.84$^*$ &   TW    \\
  40307              &        DDP     &  K3V       &  0.70 $\pm$ 0.01      & 6.0  $\pm$  4.1     &  (e)  &   4774      &   -0.36 $\pm$ 0.03  &  4.42  & 3.0        & $\leq$ 0.3              &    -4.68$^*$ &   G10    \\
  40979              &        DDP     &  F8        &  1.21 $\pm$ 0.02      & 0.8  $\pm$  0.6     & (b)   &   6145      &    0.21$\pm$  0.06  &  4.31  & 6.4        &        2.96 $\pm$ 0.11  &    -2.86$^*$ &   L06        \\
  45184              &        DDP     &  G2V       &  1.00 $\pm$ 0.01      & 4.4  $\pm$  2.7     &  (f)  &   5869      &   0.04 $\pm$  0.03  &  4.47  & 0.8        &        2.02 $\pm$ 0.12  &    -0.39$^\dagger$   &   S08.A18       \\
  50499              &        DDP     &  G1V       &  1.22 $\pm$ 0.03      & 3.8  $\pm$  0.5     &  (e)  &   5927      &   0.22 $\pm$  0.04  &   4.05 & 3.8        &        2.62 $\pm$ 0.10  &    -3.49$^*$ &   G10        \\

 \hline                                                                                                                                                                          
  \hline                                                                                                                                                                               
 \end{tabular}               -                                                                                                                                                   
 \end{table*}

 \setcounter{table}{0}                                                                                                                                                                 
 \begin{table*}                                                                                                                   
 \caption{Parameters of the stellar sample}                                                                                                                                                                     
 \label{t:parameters0}                                                                                                                                                                 
 \small
 \begin{tabular}{lcccccccccccc}
 \hline
 \hline                                                                                                                                             
  Star     &  Group   &    Sp.    &   $M_{\star}$             &  Age    & IR               & T$_{\rm eff}$    & [Fe/H] &  log ${\it g}$   &  $\mathfrak{v}\sin i$    & A(Li)  & Log($M_{\rm d}$)  & Ref\\                         
    HD      &           &    Type      &     (M$_{\odot}$)       &  (Gyr)                &    Ref.         &   (K)            &   (dex)     &     (cgs)            &(kms$^{-1}$)     &       &    ($M_{\rm moon}$)   &         \\
 \hline                                                                                                                                                        
50554              &        DDP     &  F8        &    1.10 $\pm$  0.03    & 2.1  $\pm$  1.6   &    (e)   &   6020     & 0.05 $\pm$  0.06  &  4.40    & 2.3    &           2.42 $\pm$ 0.11 &    -1.54$^\dagger$ &   TW        \\                                                                                                                                                                             
52265               &        DDP      &  G0V       &  1.20 $\pm$ 0.03   &  3.9 $\pm$  1.0   &    (g)   &   6136     & 0.21   $\pm$  0.04    &  4.36    & 3.6    &         2.65 $\pm$ 0.11 &    -1.85$^\dagger$  &    G10      \\ 
69830               &        DDP      &  K0V       &  0.95 $\pm$ 0.04   &  4.2 $\pm$  3.6   &    (a)   &   5420     & 0.02   $\pm$  0.06    &  4.35    & 0.5    &         0.88 $\pm$ 0.35 &    -1.70$^*$  &    TW      \\
73526               &        DDP      &  G6V       &  1.09 $\pm$ 0.05   &  5.0 $\pm$  2.8   &    (b)   &   5699     & 0.27   $\pm$  0.06    &   4.25   & 1.69   &  $\leq$ 0.63            &    -2.95$^*$  &    S04.DM14      \\
82943               &        DDP      &  G0        &  1.20 $\pm$ 0.02   &  1.0 $\pm$  1.0   &    (g)   &   6011     & 0.28   $\pm$  0.03    &   4.37   & 2.75   &         2.47 $\pm$ 0.10 &    -1.17$^\dagger$  &    G10      \\
108874              &        DDP      &  G5        &  0.81 $\pm$ 0.05   &  6.2 $\pm$  2.5   &    (f)   &   5572     & 0.20   $\pm$  0.06    &   4.25   & 6.7    &  $\leq$ 0.58            &    -1.96$^\dagger$  &    G10      \\
113337              &        DDP      &  F6V       &  1.40 $\pm$ 0.04   &  1.5 $\pm$  0.9   &    (f)   &   6576     & 0.13   $\pm$  0.04    &   4.21   & 2      &   --                    &    -0.72$^\dagger$  &    M12.C11      \\
115617              &        DDP      &  G5V       &  0.97 $\pm$ 0.04   &  3.5 $\pm$  3.0   &    (g)   &   5618     & 0.02   $\pm$  0.03    &   4.52   & 8.8    &  $\leq$ 0.22            &    -2.24$^\dagger$  &    G10      \\
117176              &        DDP      &  G5V       &  1.06 $\pm$ 0.03   &  8.1 $\pm$  0.4   &    (b)   &   5560     & -0.06  $\pm$  0.03    &   4.07   & 0.8    &         1.85            &    -1.64$^\dagger$  &    L06.GO10.R12      \\
128311              &        DDP      &  K0        &  0.81 $\pm$ 0.04   &  1.0 $\pm$  0.5   &    (f)   &   5120     & 0.01   $\pm$  0.03    &   4.49   & 1.36   &  $\leq$ 0.13            &    -3.91$^\dagger$  &    G10      \\
130322              &        DDP      &  K0V       &  0.90 $\pm$ 0.03   &  4.1 $\pm$  3.6   &    (e)   &   5373     & 0.01   $\pm$  0.04    &   4.33   & 3.2    &  $\leq$ 0.38            &    -3.10$^*$  &    G10      \\
150706              &        DDP      &  G0        &  1.03 $\pm$ 0.02   &  0.3 $\pm$  0.3   &    (b)   &   5961     & -0.01  $\pm$  0.05    &   4.50   & 1.8    &         2.63            &     --  &    GO10.R12      \\
178911B             &        DDP      &  G5        &  1.02 $\pm$ 0.03   &  4.2 $\pm$  1.8   &    (b)   &   5600     &  0.27  $\pm$  0.04    &   4.44   &  --    &  $\leq$ 0.47            &    --     &    L06.R12     \\
187085              &        DDP      &  G0V       &  1.24 $\pm$ 0.03   &  1.2 $\pm$  0.9   &    (e)   &   6190     & 0.21   $\pm$  0.06    &   4.45   &  --    &         2.61 $\pm$ 0.11 &     --    &    TW     \\
192263              &        DDP      &  K0        &  0.80 $\pm$ 0.05   &  3.4 $\pm$  3.4   &    (b)   &   5028     & -0.06  $\pm$  0.05    &   4.29   & 3.74   &  $\leq$ 0.02            &    -3.46$^\dagger$  &    G10      \\
210277              &        DDP      &  G0        &  0.95 $\pm$ 0.02   &  5.0 $\pm$  3.3   &    (o)   &   5505     & 0.18   $\pm$  0.04    &   4.30   & 5.7    &  $\leq$ 0.41            &    -2.50$^*$  &    G10      \\
215152              &        DDP      &  K0        &  0.76 $\pm$ 0.07   &  5.2 $\pm$  4.0   &    (i)   &   4803     & -0.08  $\pm$  0.09    &   4.26   & 1.8    &          --             &    -4.39$^*$  &    S08      \\
216435              &        DDP      &  G3IV      &  1.30 $\pm$ 0.03   &  3.4 $\pm$  0.5   &    (g)   &   5993     & 0.20   $\pm$  0.07   &   4.14    & 3.35   &         2.65 $\pm$ 0.11 &    -1.96$^*$  &    TW     \\
224693              &        DDP      &  G2V       &  1.31 $\pm$ 0.09   &  3.4               &    (e)   &   5960    & 0.28  $\pm$  0.07    &   4.06   & 3.7    &          2.03 $\pm$ .13  &     --     &    TW      \\
\hline
\hline 
\end{tabular}                                                                                                             
{ Col.1: HD star name; Col.2:  Group defined in section 1: C, CP., DD and DDP;                                                                           
Col.3: spectral type; Col.4 and 5: stellar mass and age, respectively; Col.6: references of IR data: References: \cite{trilling08} (a), \cite{kospal09} (b),  \cite{plavchan09} (c),
 \cite{bryden06} (d),  \cite{bryden09} (e), \cite{chen2014} (f), \cite{maldonado12} (g),     \cite{beichman06b} (h),
\cite{koerner10} (i), \cite{greaves04} (j), \cite{hillenbrand08} (k), \cite{rhee07} (l), \cite{kalas06} (m),  \cite{carpenter08} (n),  \cite{eiroa2013} (o); Col.7: effective temperature; Col. 8: metallicity; Col. 9:  surface gravity; Col. 10: projected rotational velocity;  Col. 11: lithium abundance; Col. 12: log of mass of the dusty disc and references, where the symbol ``$\dagger$''  means taken from \cite{chen2014},    and the symbol ``$*$'' means calculated in this work as explained in section 3.0; Col 13: References of stellar parameters and A(li); TW, This work; G10, \cite{ghezzi2010b}; GO10, \cite{gonzalez2010}; S08, \cite{sousa2008}; L06, \cite{luckHeiter2006}; R12, \cite{ramirez2012}; M12, \cite{maldonado12}; S04, \cite{santos04}; DM14, \cite{delgadomena2014};  V05, \cite{valenti05}; B91, \cite{boesgaard1991}; C11, \cite{casagrande2011}; E12, \cite{ertel2012}; G16, \cite{gaspar2016}; S06, \cite{sousa06}; L17, \cite{luck2017};    A18, \cite{aguilera2018}; B15, \cite{bell2015}}
\end{table*}

\section{Metallicity properties} \label{mdisc}

In the Core-Accretion model for planetary formation (hereafter CA) it is expected that cores of giant planets will be  preferentially formed in a
high metallicity medium, where solid elements are abundant. 

 Stars with only debris discs do not show metal enrichment. As planetesimals are the
building blocks of planets, and gas-giant planet hosts show the metal signature,
this constitutes an apparent paradox that might be raised as an argument against
the CA model  \citep{fletcher2016}. Furthermore,  we can ask why there are plenty of high metal abundant stars hosting debris disc without giant planets?

 \cite{greaves2006} consider that the  lack of a metallicity correlation 
 of DD stars and also, the known giant planet - metallicity relation
 are both situations, in agreement with the CA model \citep[see for instance ][]{fischeryvalenti05}. For \cite{greaves2006} the effect of higher solid metal
abundances, for a given PP disc mass, will speed giant planet formation in a gas rich PP disc, before the gas vanishes in less than around 10 Myr.
For these authors, the non-metal behaviour of DD stars, will arise due to a different metallic dependence, in which planetary cores without gas could be
formed at later epochs than the short lived PP disc phase. \cite{Wyatt2007}  studied the origin of the metallicity dependence
on the PP disc of a star with planets. However, any prediction appears to be made only for A-type stars.

During those years, the discussions on metallicity were made by considering mainly two different stellar cases; those of stars with gaseous
giant planets mainly (CP) and those of DD stars, i.e., without giant planets. One important ingredient for a complete discussion on
this subject consists to introduce the case of DD stars containing giant gaseous planets (DDP). Few authors as \cite{maldonado12,maldonado15}
have considered this case. In this paper  we  investigate more deeply these aspects on metallicity and also on lithium by means of a  tool
consisting in using the masses of the dusty debris discs (hereafter $M_{\rm d}$) in stars with and without planets. For this purpose, we have collected the
largest number of DDP stars known in the literature and found   around 30 main-sequence objects.
This number fixed in a certain way, the numbers of the other three selected groups: 41 DD stars, 31 CP stars  and  38 C stars. 
A relatively similar number of objects will be then considered for each group for better comparison purposes. 
Moreover, we have not considered very close binary stars and nor any star in the sub-giant evolutionary stage to avoid introducing a bias
in our analysis. This is  because previous studies indicate that there are differences in both disc frequency and planet
frequency between single and binary stars \citep{rodriguezzuckerman},  and also a  different behaviour in respect with metallicity for planet-hosting giant stars.
Regarding the metallicity distribution of evolved stars
with planets there are many  works showing puzzling results: \cite{pasquini07,sadakane2005,schuler2005,hekkerMelendez2007,takeda2008,Ghezzi2010GIANT,maldonado2013,mortier2013,jofre2015,reffert2015,maldonado2016}.

In this analysis, the values of the disc masses $M_{\rm d}$ were mainly obtained from \cite{chen2014}.
This catalog present two types of discs for a given star. A small and warm disc, and a large and cold disc. For coherence
purposes, we select in this work always the largest discs corresponding to the largest $M_{\rm d}$ values for each object. In the case of the \cite{chen2014}.
catalog the value is that of M$_{\rm d2}$ 
(second dust mass in the two -T model).  For DD or DDP stars not contained in this mentioned catalog, we estimate the corresponding
$M_{\rm d}$ values by using equation number eleven of \cite{chen2014} work. The use of this equation requires, apart the observed fractional infrared luminosity
(L$_{\rm IR}$/L$_{*}$) ratios, the appropriate discs radii which have been obtained from the Spanish Virtual Observatory  (SVO) catalog (http://svo2.cab.inta-csic.es/vocats/debris2/). These calculations require specific values of the grains radii for the spectral type of the star in consideration. These values are taken from the table 3 of \cite{chen2014}. As chosen in that work, the mass density is equal to 3.3 gcm$^{-3}$. These estimated  $M_{\rm d}$ values (in lunar mass units) are indicated with the symbol  ``$*$'' (asteristic) in Table \ref{t:parameters0}.
The main assumptions of the models of debris discs in \cite{chen2014} are that an important part of the dusty discs can be better described by two,
warm and cool, dust components with two temperature blackbody models. Also, no collisional models have been applied by these authors to
determine the disc masses. In our case, we warn that we have chosen, for homogeneity comparative purposes, the largest cool discs for all our stars
with discs. This means that our selection is independent if real discs have multiple components or not \citep[see also][]{KennedyWyatt2014}.

\subsection{ Metallicity of debris discs without giant planets (DD)}

\begin{figure*}
   \includegraphics[width=0.9\columnwidth]{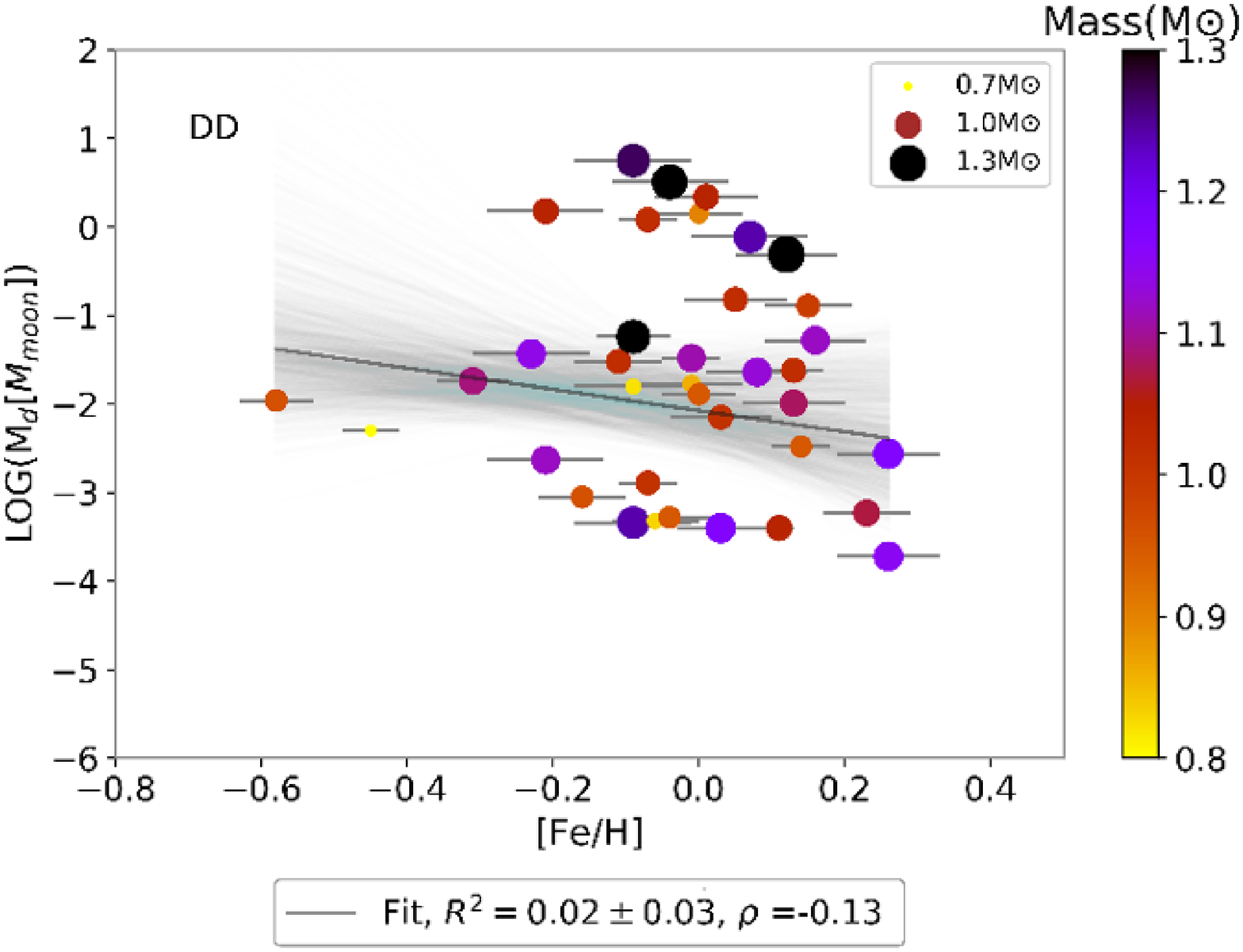}
   \includegraphics[width=0.9\columnwidth]{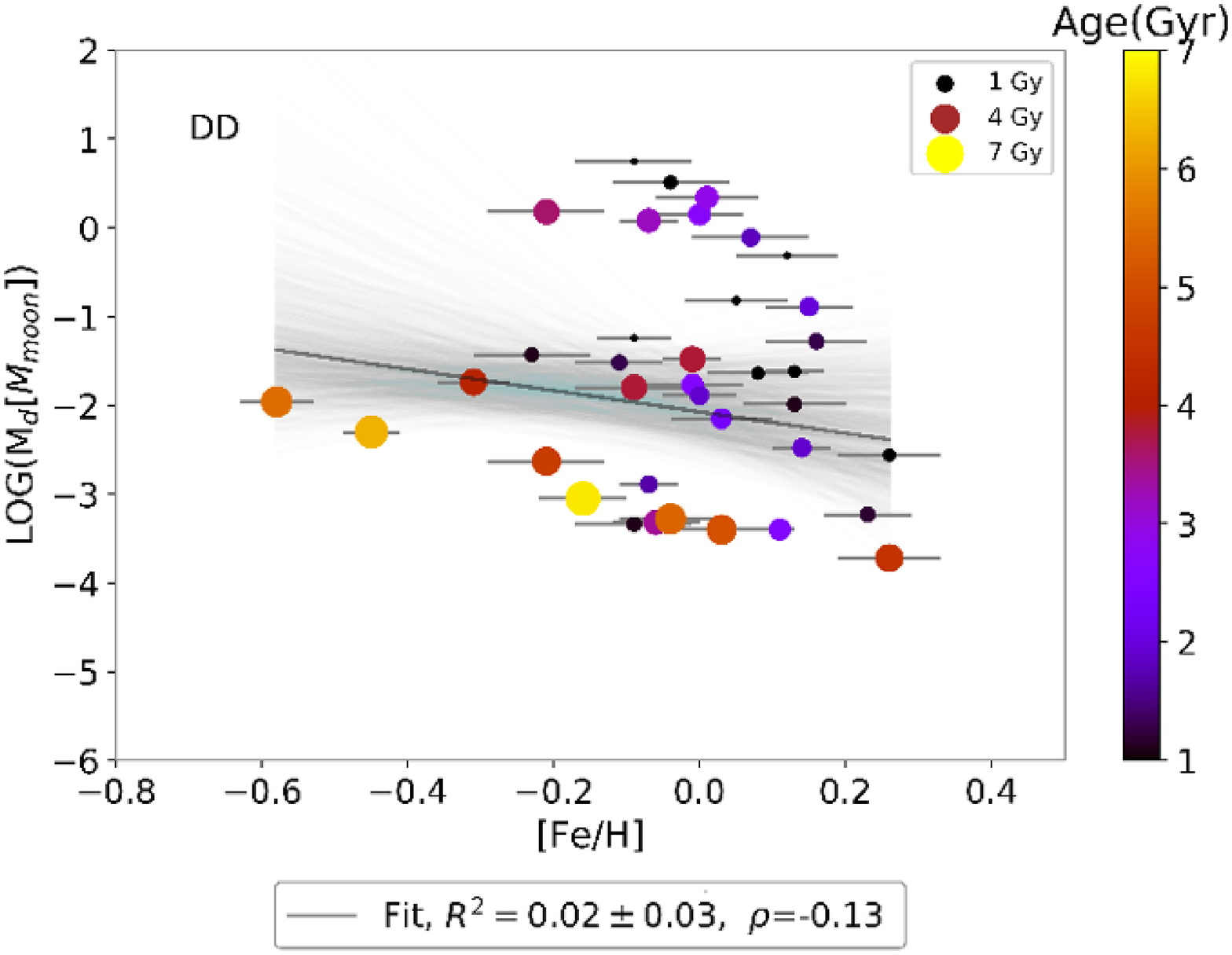} 
 \caption{ Stellar metallicity as function of the mass of the dusty disc, segregating with stellar mass (left panel) and age (right panel) for the DD stars. The solid black line shows the  best lineal regression   (y = ax + b), calculated using
bootstrapping method weighted by the errors of metallicity, the parameter values are shown in the box below 
the figure. The shaded blue area shows the 68\% confidence  band of the  bootstrapping fit.
 Note that the slope attains  positive and negative values, which is reflected also in the values of $R^2=0.02$ and  $\rho$ $=$ -0.13, which mean not correlation.
}
   \label{DDMdMetal}
    \end{figure*}

As pointed out before, we called   the DD  group  to the stars with dusty discs without containing observed planets of any mass. 
Figure \ref{DDMdMetal} shows  the main result of the behaviour of the  masses of the debris disc, $M_{\rm d}$,  in function of metallicities of the DD 
host stars. The absence of correlation is evident by the low statistical correlation values showed in the figure (Pearson coefficient $R^2=0.02$). 
The linear regressions are performed by applying a bootstrap technique using Monte Carlo (MC) method.  First, we generate 1000 bootstrapped samples from the original sample. Each bootstrap is a random draw with replacement from the original sample with a draw size equal to the original sample size. Second,
we calculate the lineal regression of each bootstrapped sample. The goodness of the fittings are shown with the shaded
areas (68 per cent confidence bands). The shaded blue area is formed for the 1000 linear regression fits   calculated by the boostrap method, in which  it is easy to see the slopes taking positive, null and negative values. 
Additionally, we applied the Spearman rank correlation test, widely used in astronomy to discern whether a set of two variables are correlated or not \citep[see e.g.][among others]{damasso2015,koljonen2015,patruno2016}. Spearman's rank correlation coefficient is denoted by the symbol $\rho$,  where  $\rho$ $= 0$  corresponds to no correlation between the variables, while   $\rho=+1$ or $\rho=-1$ corresponds to a perfect increasing or decreasing monotonic correlation.
We implemented the code described in \cite{curran2014}, obtaining $\rho$ $=$ -0.14. Both tests, using Pearson  or Spearmann coefficient, show  null or very low correlation, respectively,  between stellar metallicity and the mass of the debris disc for the DD sample.
This result is not a surprise and agrees with past investigations on DD on the metallicity, or any other characteristics
\citep[e.g. ][]{beichman05a,chavero06,greaves2006,moro-martin07b,bryden09,kospal09}. 

As far as we know, the only investigation using $M_{\rm d}$ values to search for a correlation of DD with [Fe/H] is that of \cite{gaspar2016}. However,
they  propose to have found a correlation with metallicity, only for the ``de-aged'' $M_{\rm d}$ values calculated by them from the 
actual age $M_{\rm d}$ values. These ``de-aged''  values corresponds to an age of 1 Myr, this is near the initial age of the PP disc. Nevertheless, as seen in their
figure 7 the claimed correlation seems to appear for sub-solar metallicities only. Gaspar et al. used in the search for correlations the totality of DD stars of their 
catalog. This totality contains a mixture of  upper limit flux values, binaries or multiple objects and  also stars in different evolution stage and spectral types.
Also, they do not made any distinction between DD and DDP objects.

In addition, \cite{gaspar2016}  show an important absence of highly metal depleted DD objects. However, this deficit of metal poor DD especially in the
interval (i.e., $-$0.50 $<$ [Fe/H] $<$ $-$0.20) were already discussed in \cite{montesinos2016} and references therein. This deficit  is in a strong contrast with 
the relative large number of DD objects in the enhanced metal region. In any case, we confirm  these mentioned  deficit results, by means of the asymmetrical 
distribution of our DD objects shown in Figure \ref{DDMdMetal}.
In principle, this absence of DD stars in very low metal regimes, can be explained by an evolution age effect. In fact, the mean age founded of DD stars is 2.5 Gyr, being the youngest group among the four groups studied.
In general, these young DD stars are represented by the large population of metal enriched DD stars. On the contrary, among the most depleted DD stars we found the oldest DD objects. That is the case, for instance, of HD 110897 with 
[Fe/H] $=$ $-$0.59 and an age of 5.52 Gyr and HD 158633 with [Fe/H] $=$ $-$0.45 with an  age of 6.3 Gyr, we can then say that there are few DD survivor stars
in very low metal regimes, because similar objects as those mentioned, had already remove their dusty debris content.

We can conclude that the amount of  dust in DD stars  does not  depend of the stellar metallicity.  It must be remembered that in general, the disc masses
presented in this analysis do not represent the real masses of the discs.
This is  because in all cases, any large km-sized bodies as planetesimals, are not directly observable. We can only measure the masses of the dust (MIR) and of the planets. The real total mass of the disc may be higher.

\subsection{Metallicity of debris discs with planets  (DDP)}

Stars containing dusty debris discs and planets (DDP) are surely one of the best laboratories to test the CA planetary formation theory  \citep{pollack96}.
However, differently than the other groups considered in this work; DD, CP and C  which are more numerous, the total number of DDP  stars known today, is around 33
(data taken from SVO catalog, http://svo2.cab.inta-csic.es/vocats/debris2/).
 This number has only  increased slightly since the 22 DDPs objects proposed by \cite{kospal09}. 
We must note that the increasing  in the  number of  stars with  discs and  planets with respect to the \cite{kospal09}   work is mainly due to the discovery of low-mass planets around debris disc stars. In fact, using the exoplanet 
catalog \citep{ schneider2011} we learn that before the year 2010, the mean planetary mass detected for DDP stars was of 1.76$M_{\rm jup}$. After, in the period of 
years 2010-2018, this detected mean planetary mass attained a mean value of 
0.03 $M_{\rm jup}$.

Eliminating all close binary stars \citep{raghavan2006} our final list of main-sequence DDP objects contains 30 stars. From this number only ten  have their discs spatially resolved.
The majority of the selected DDP stars contain close-in planets with debris discs at large radial separation. Some details on the planets-disc interaction
have been discussed by \cite{hughes2018} and \cite{wyatt2018}. A recent important direct imaging survey of Spitzer sources, searching for the
presence of more massive  gaseous planets ($>$ 5 $M_{\rm jup}$) in debris discs stars, produced no new discoveries
\citep{meshkat2017}.

\subsubsection{ The stellar case: metallicity vs stellar mass }

\begin{figure*}
\includegraphics[width=0.8\columnwidth]{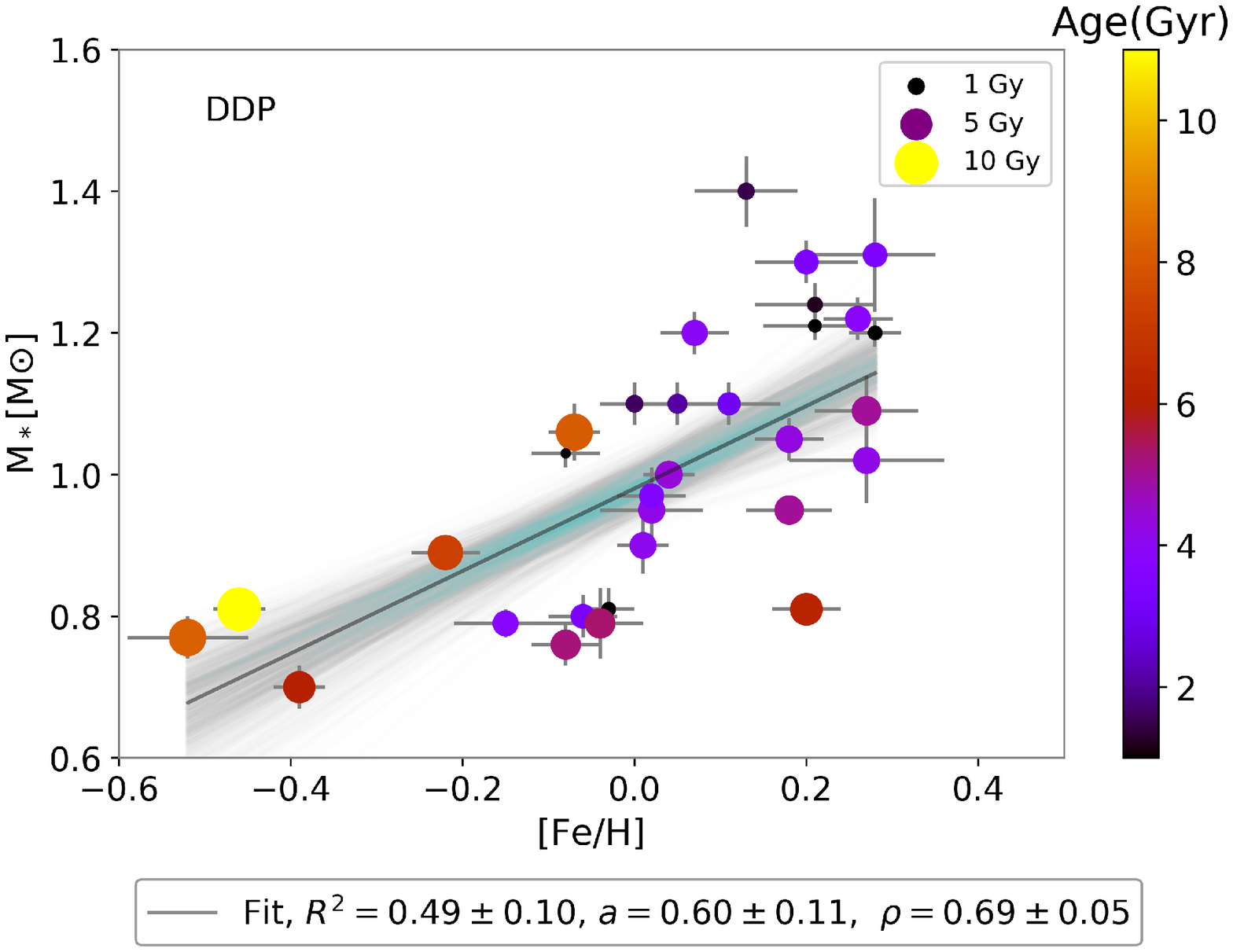}
\includegraphics[width=0.8\columnwidth]{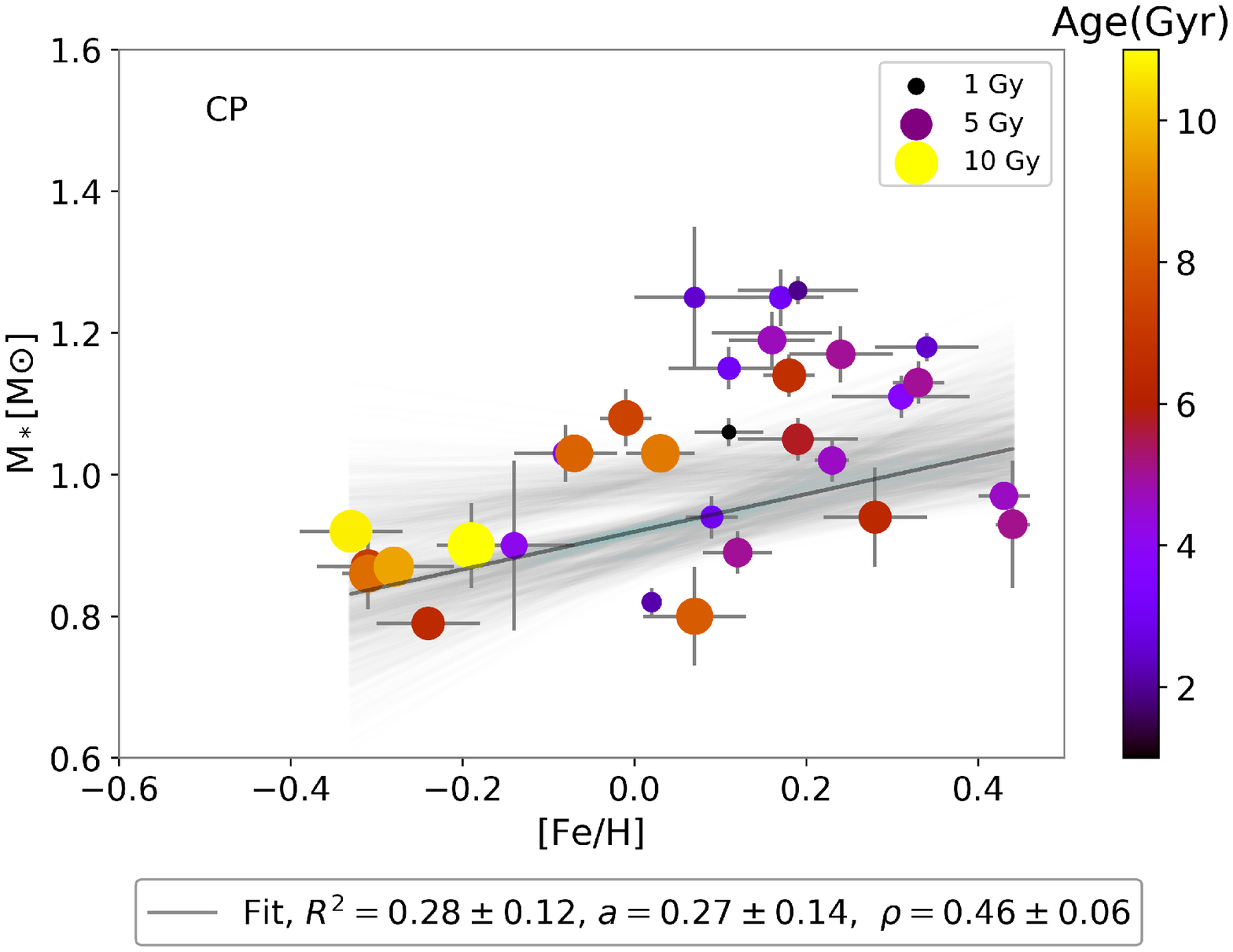}
\includegraphics[width=0.8\columnwidth]{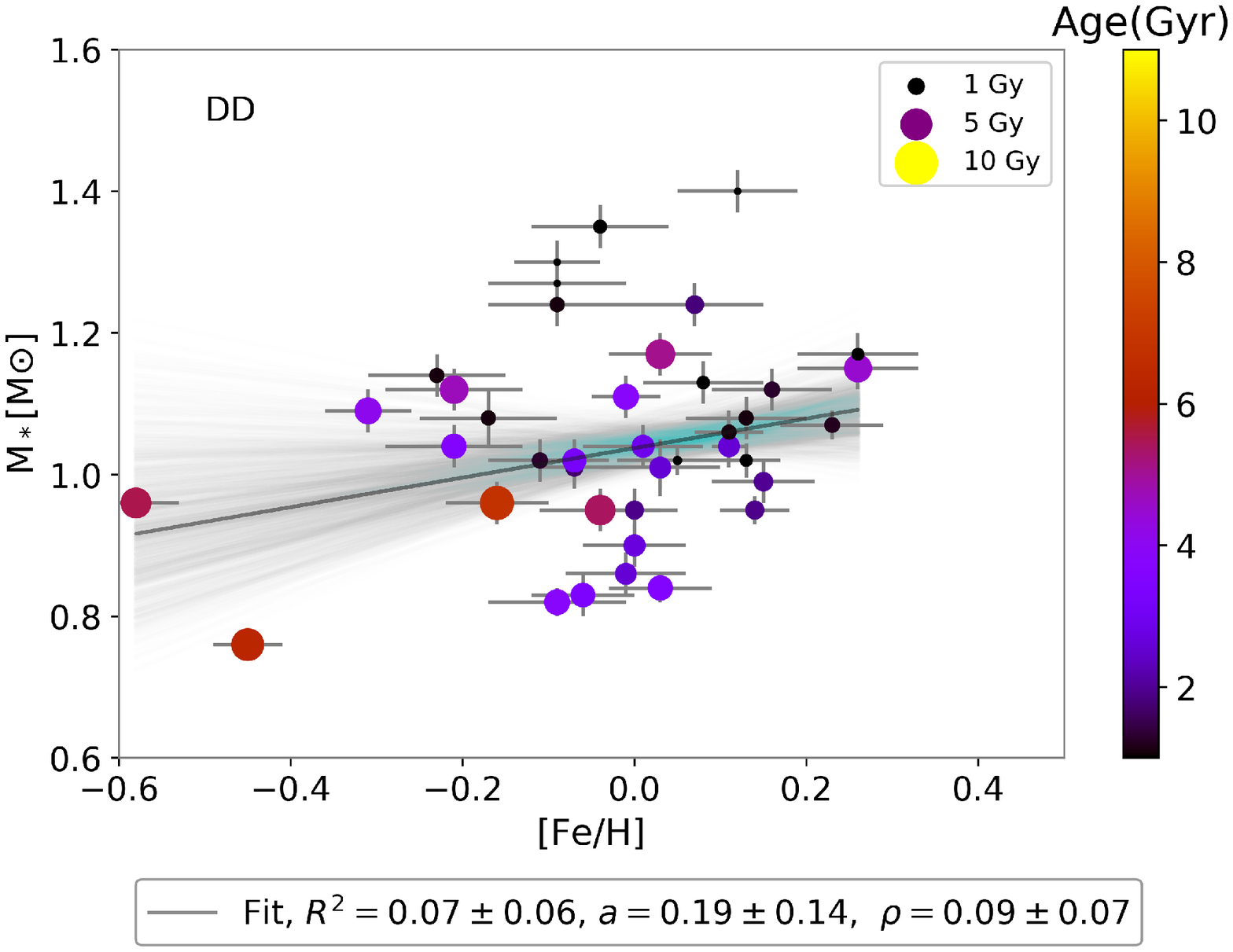}
\includegraphics[width=0.8\columnwidth]{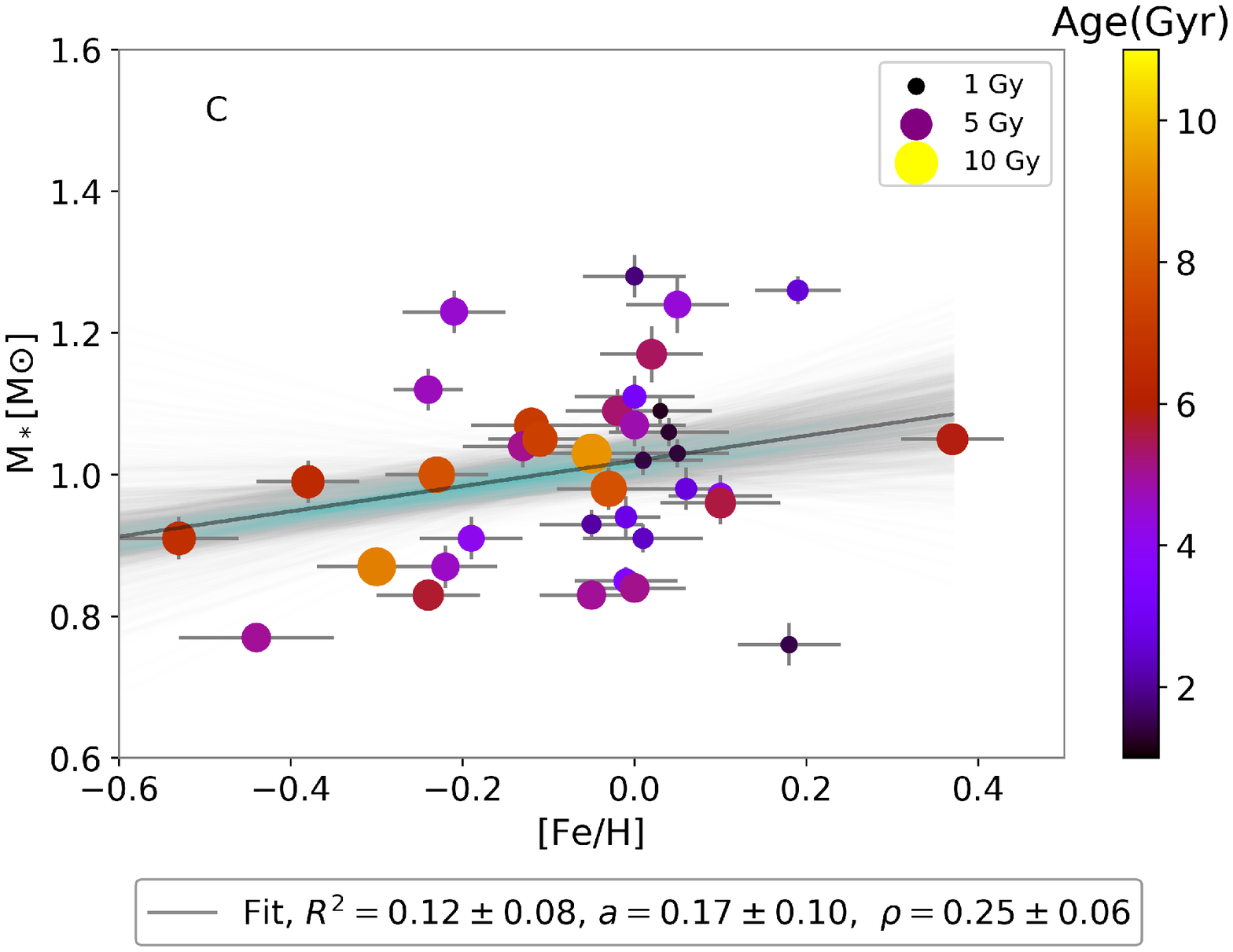}

\caption{ Stellar mass  as function of star metallicities for the four samples, segregating with age in the color scale bar for the four groups. Each panel showsthe best linear fit in solid line (y = ax + b) and  the values of the parameters, calculated using bootstrapping method weighted by the errors of metallicity. The shaded  area shows the 68\% confidence interval  on the mean from bootstrapping (1 $\sigma$). The upper panels show an clear increase of the stellar mass with  the host star metallicity, being that  DDP stars present the steepest slope  with the best 
coefficient of determination R-squared and $\rho$. In the case of stars without planets, (bottom panels) the slope decreases. Particularly, in the case of DD stars, the very low R-squared coefficient  reveals no relation at all.
}
\label{massmetal}
    \end{figure*}

The metallicity of DDP stars has been studied by \cite{maldonado12,maldonado15} considering the same definition of the  four groups as in this work. Even if the number of
objects of groups DD, CP and C are nearly four times larger than ours, their number of DDP stars is similar to  our analysis. These authors found
by means of K-S statistical distributions, similar metallicity properties for  DDP and CP groups. These two being however, different of those of DD and C groups.
They claim that the mentioned similarity of DDP and CP groups is due to the presence of planets only and not due to the debris. Differently from
 \cite{maldonado12,maldonado15}, we use  approximately the same number of members for the four groups
 to investigate the metallicity dependence using a different approach based mainly on the use of stellar and dust
discs masses.  Figure \ref{massmetal} shows the distribution of stellar masses in function of metallicity for the four groups considered in this work.

As in Section 3.1, we have  performed a search for correlations, this time between the stellar  mass and the metallicity
for the four groups, using  the classical Pearson and the Spearman correlation tests presented before, but in this case taking into account the errors in both parameters. 
 
 DDP and CP stars present the larger rate of increase of the stellar mass with metallicity, being remarkable that DDP stars present a slope value
twice of the  CP stars.  Also, these two groups present good correlate  fit values, being $R^2= 0.28\pm0.12$  ($\rho=0.46\pm0.06$) for CP stars and  $R^2= 0.49\pm0.10$ ($\rho=0.69\pm0.05$) for DDP stars.
The C group of stars show a poor correlation, whereas DD stars show an absence of correlation, characteristic of their disorder state as
discussed before concerning their dusty disc masses. We obtain then an
important conclusion for DD stars; both their stellar masses and discs appear 
to be constructed in  complete disorder in respect to metallicity.

We can conclude that it is  
the presence of planets that determine the increasing orderly
steps in function of metallicity. We note, however, that 
similar positive increases of the stellar mass (not for DD stars) with metallicity have been presented by  \cite{gonzalez06b} 
 \citep[see also][]{ghezzi2010a}. The common interpretation of this behaviour is
due to the Galactic stellar age-metallicity relation, where young stars are more metal rich and
at the same time, more massive stars have shorter main-sequence lifetimes.

In respect to the presence of giant planets,  Figure  \ref{massmetal}
 reveals a tendency of higher  stellar masses  of DDP and CP stars with decreasing age.
  It is important to note that even  the mean stellar masses of the four stellar
groups, are practically, all them similar to 1.0$M_{\odot}$, their distributions on a
histogram in function of stellar masses are differents. Whereas for groups
without giant planets as C and DD, their histogram distributions are peaked at
$\sim$ 1.0 $M_{\odot}$, contrary, the groups with planets CP and DDP have a flat
distribution between 0.7 and 1.3 $M_{\odot}$. These differences of distributions are
interesting and reveal, when transformed into metallicities, the different
gradients of the groups contained in Figure \ref{massmetal}.
These distributions also explain why stars, in their quest to have
larger metallicities to form giant planets more efficiently, they need masses
around 1.2 -- 1.3 $M_{\odot}$  in CP and specially in DDP stars. As mentioned
before, this is due to the fact that higher masses stars, are younger and are more
metallic. These different behaviours of stars with planets and without planets
do not necessarily introduce a bias. On the contrary, it helps to understand
more the  planetary formation.

 Concerning the comparison of the metallicity distributions between groups C and CP,
we find  that CP stars are more metallic by an approximately factor of 0.15 dex.  This
already known result in the literature can be found, for instance, in  \cite{ghezzi2010a} 
where a similar metallicity shift between CP and C stars has been found. This
known result represent one of the first evidences of the stellar-planetary relation
 \citep{gonzalez97,santos04,fischeryvalenti05}. In respect to our present results for DD and DDP objects we
must note that the stepper slope for DDP stars respect to CP appears to be a new
result.

\subsubsection{ The debris disc dust case:  metallicity vs $M_{\rm d}$   }

\begin{figure}
     \centering
   \includegraphics[width=7.5cm]{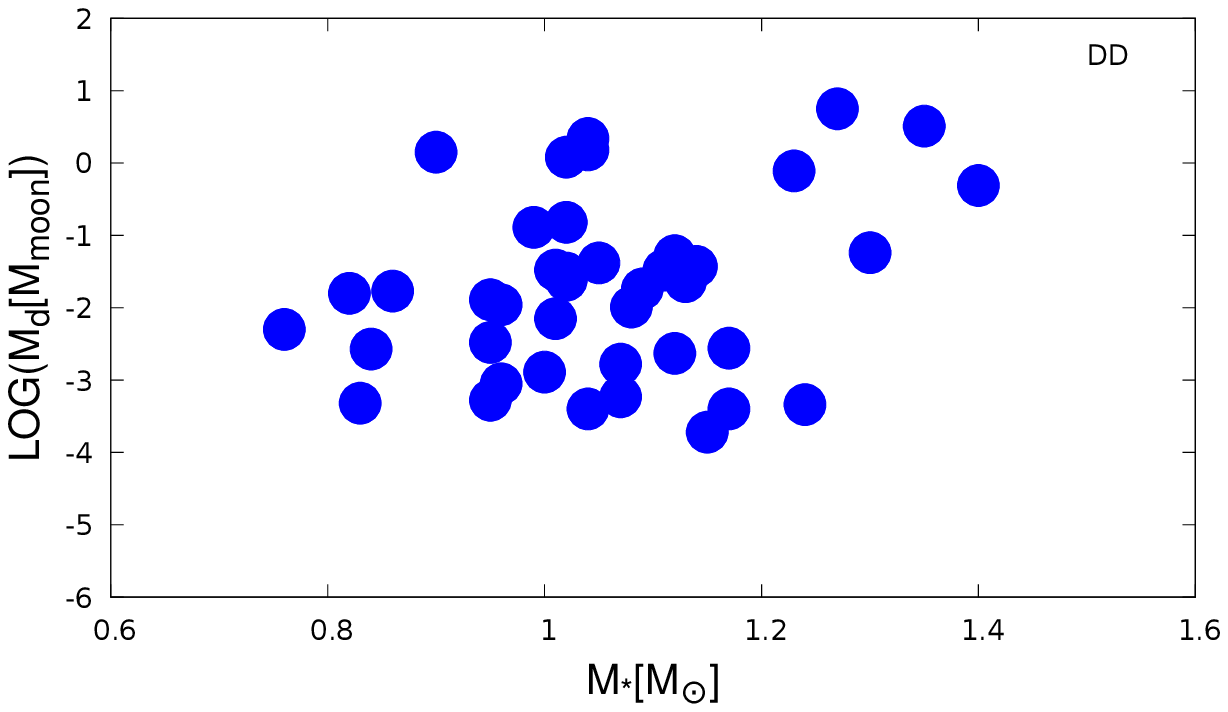}\par 
   \includegraphics[width=7.5cm]{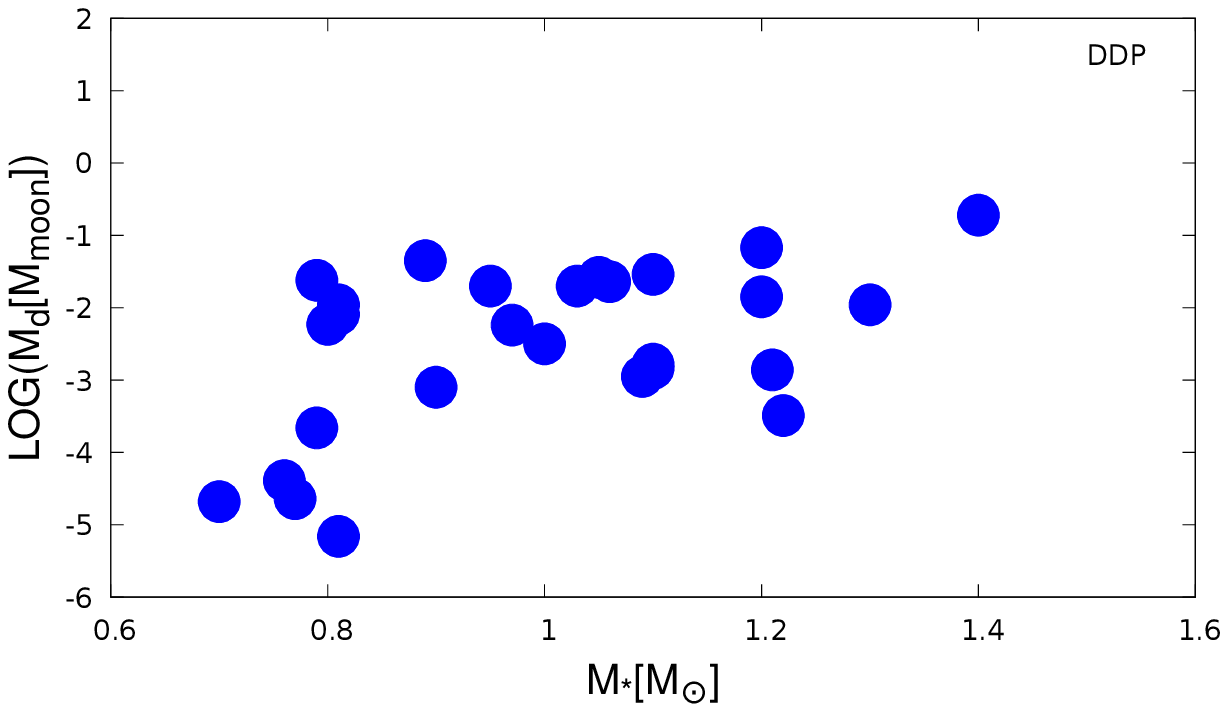}\par 
       \caption{ a) Top panel: log $M_{\rm d}$  vs. M$_{*}$   for the DD group, (b) Bottom panel:
log $M_{\rm d}$ vs. M$_{*}$  for the DDP group.  See the text for an interpretation of these distributions.   }
   \label{DD}
    \end{figure}

Now, remains the question of what is the behaviour of the dust component of DDP stars in function of the host star metallicity.
For  this purpose, we use the masses  $M_{\rm d}$ of the 
respective dust discs in function of the central stellar metallicity. To our knowledge, this approach have never been made before for the DDP systems. 
Before presenting the $M_{\rm d}$ behaviour as a function of metallicity, we present a  result which we found connecting the DDP dust masses 
with the stellar masses. This is shown in Figure \ref{DD}. This kind of $M_{\rm d}$ - M$_{*}$ relation is quite known in very young stars, in general with ages less
than 10 Myr \citep{andrews2013,pascucci2016}.

\begin{figure*}
 \includegraphics[width=0.9\columnwidth]{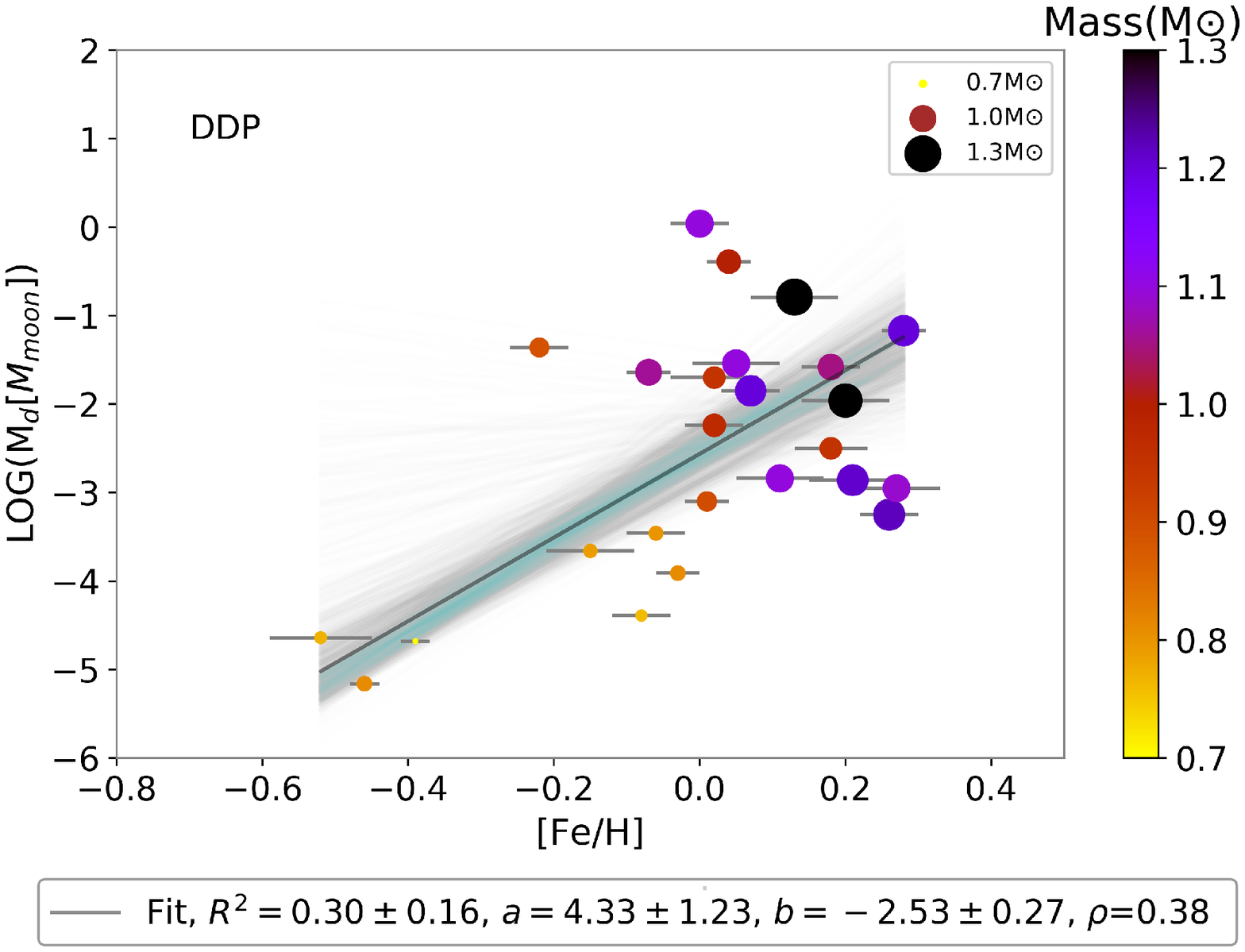}
 \includegraphics[width=0.9\columnwidth]{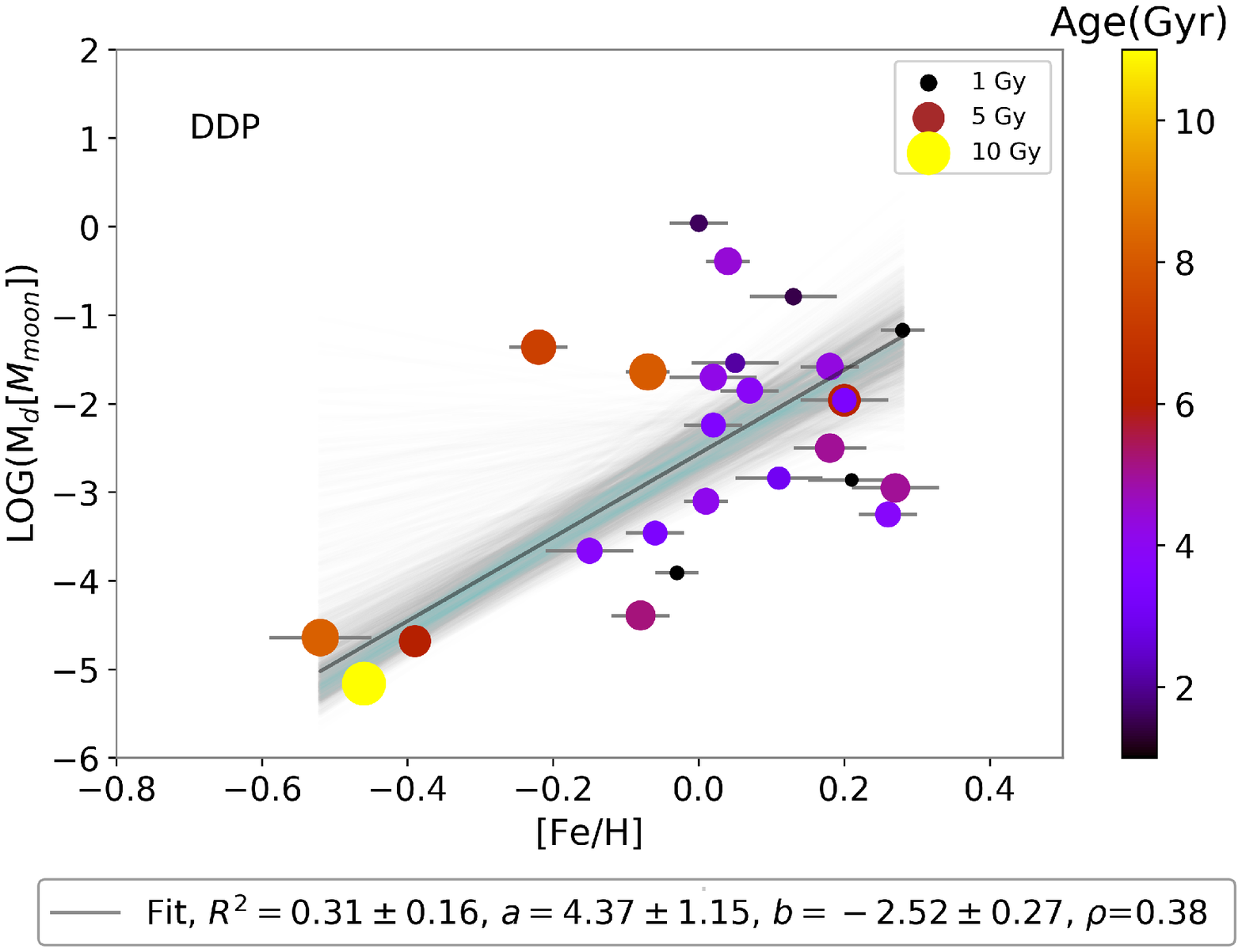}
 
\caption{Stellar metallicity as function of the mass of the dusty disc, segregating with stellar mass (left panel) and age (right panel) for DDP stars.  The solid black line shows the lineal regression (y$=$ax+b), the parameter values are shown in the box below the figure.
The shaded area shows the 68\% confidence  band of the  bootstrapping fit (1 $\sigma$).
Note that the slope attains always positive values  (even with 3 $\sigma$) and with a good of $R^2=0.31$ and  $\rho$ $=0.38$, showing a very good correlation.
}
   \label{DDPMdMetal}
    \end{figure*}

The relation of $M_{\rm d}$ for DDP stars with metallicity are the ones presented in Figure \ref{DDPMdMetal}, indicating for the first time,
a clear tendency for a correlation of these dusty discs with metallicity.  In this case, we also apply the bootstrap (MC) and Spearman methods presented before.
The DDP sample shows a much better positive correlation ($R^2=0.31$,  $\rho$=0.38) than  the DD one. Also, the slopes of the linear regression are always positives.

This result was possible, not only by the use of debris disc masses
(M$_{\rm d}$) but also by the use of the largest possible list of clean DDP objects, for which we have used objects containing no limit fluxes
of IR radiation and as mention before, not including close binary stars and retaining only  FGK main sequence stars.
Thus, only in  DDP stars  there is a systematic increase of $M_{\rm d}$ in function of  the 
stellar metallicity. This increase represents several orders of magnitude from the very low metal regime at [Fe/H] $= -0.52$ to the maximum observable at high metallicity at [Fe/H] $= +0.3$.
In Section 3.1, we presented for DD stars the Figure \ref{DDMdMetal} which is the corresponding one of Figure \ref{DDPMdMetal} for DDP stars.
Differently than DDP stars, in the DD group we found no correlation with metallicity.

\subsubsection{The planetary case: metallicity vs planetary mass }\label{planetarycase}

In the two precedent subsections we saw that both, the stellar masses and disc dust masses correlate. Also, independently, we saw that 
both masses increase with metallicity. Now, we present a third mass increasing with metallicity. This one concerns the total mass of the planets.
One of the  first indication of the existence of a
gradient of planetary masses and the metallicity of their central stars was claimed by 
 \cite{sousa2008}. The first evidence being that jupiter mass planets appear to be
found around metal rich stars. In contrast, Neptune-mass planets have been found to have
a relatively flat metallicity distribution \citep{udry06}. \cite{ghezzi2010a} revised and confirmed these
results for typical FGK type stars avoiding the problematic M-type stars which were
included in \cite{sousa2008} work.

\begin{figure}
 \centering
  \includegraphics[width=7.5cm]{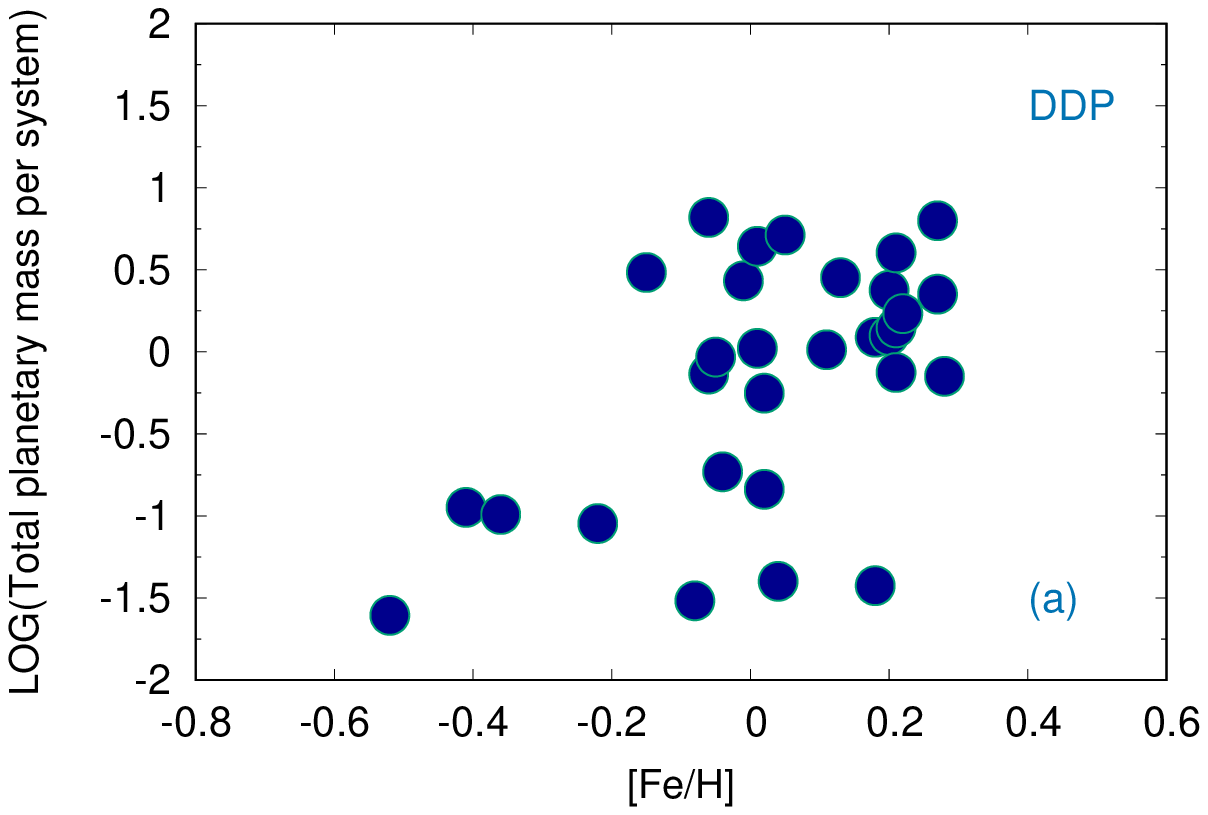}\par  
    \includegraphics[width=7.5cm]{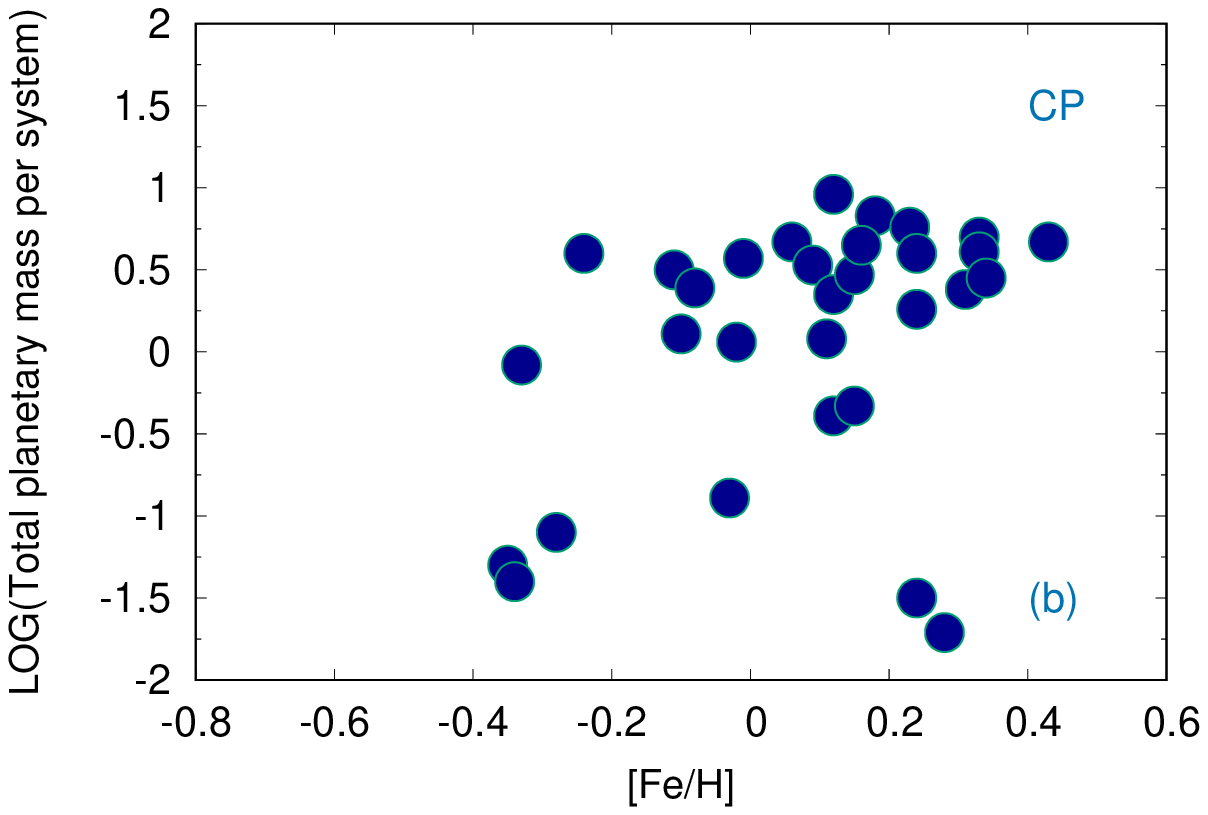}\par

 \caption{ Log of the total planetary mass per system (in Jupiter mass unit) vs the metallicity of
the star. Panel a) DDP sample; panel b) CP sample. The corresponding
Spearman correlation coefficients are $\rho=0.36$ for CP stars (moderate
correlation) and $\rho=0.18$ (less than moderate correlation) for DDP stars.
Even if the most massive planets indicate an approximate general
positive relation between M$_{\rm planet}$ and [Fe/H], the normal presence of minor
planets indicates that these relations cannot be completely linear. Also,
the fact that the correlation coefficient is smaller for DDP stars than the
one for CP stars, is compatible with the scenario that the presence of
stable dust discs are favourable to the formation of low-mass planets. 
  }
   \label{planets-iron}
    \end{figure}

Now, we consider the total mass of the known detected planets around a given star.
All planets considered around  DDP and CP stars are presented in Table \ref{t:planetsDDP} and \ref{t:planetsCP}, respectively.
In order to avoid the inclusion of brown dwarfs we limit the
  planet masses ($M_{\rm planet}$) to a maximum of 13 $\pm$ 0.8 $M_{\rm jup}$ \citep{spiegel2011}. 
The more massive planet of our sample correponds to HD 33564 b with 9.1 $M_{\rm jup}$ (CP sample).
We present in Figure \ref{planets-iron} the distribution of the total planetary masses for each star in function of the metallicity of the respective stars. 
Rigorously this is not a direct relation of the total masses of the planets with [Fe/H],
but only with the solid cores masses inside the planets. This is because a large part of
the planetary masses is due to their H and He gas component and  does not depend
 on metallicity. This relation could then better be called a ``cores of planets -- metallicity relation'' (see Fig \ref{planets-iron}).

As pointed out before, two main different properties are related to the debris disc masses
($M_{\rm d}$). One the one hand, they increase with the metallicity of the central star and
independently, on the other hand, they increase with the mass of the central star. All
these properties leave us to make the following considerations: because at least, giant
gas planets observed in DDP stars, were formed during the protoplanetary stage, we can propose
that two main conditions are necessary to form these giant gaseous planets: 1) a
sufficient larger metallicity and 2) a sufficient larger total (gas and dust) mass of the
PP disc. Nevertheless, we note that a critical minimum mass must exist in order to
form giant planets. If the PP disc mass is always less than this critical mass, no giant
planets can be formed and this could be the case of DD stars.

Also, we can consider that both mentioned conditions can be somewhat complementary.  For example, in the
case of the presence of a giant planet in a metal deficient star, we can infer that the total mass of its protoplanetary disc  was exceptionally large in order to create a giant planet in a low metallicity regime \citep[see also][]{alibert2011,mordasini2012,ghezzi2018}.

This reasoning can also be made in the opposite direction.
Differently to the disorder (i.e., lack of correlation) appearing in DD stars with respect to metallicity,  in
the DDP case, an order has been created. Then, each point in the ascending tendency
of Figure \ref{DDPMdMetal}, represents not only a given larger metallicity, but also a given larger
total mass of its dusty disc.   Also, by changing the 3D parameter representing the mass in Figure \ref{DDPMdMetal}-a to age in Figure \ref{DDPMdMetal}-b, it is noticeable that the most massive systems (stars, dust and planets) are the younger ones.

It is known that dust disc masses diminish with age. This property can
clearly be seen in the right panel of Figure \ref{DDPMdMetal} where we can observe that the
most massive dusty discs are the younger ones and the least massive are
the oldest ones. This is due to the fact that as age increases, the cold dust
(chosen by us) mean temperature in the two belt system decreases.
Collecting this property with the similar stellar mass behaviours (Figure \ref{massmetal})
and those of planetary masses (Figure \ref{planets-iron}), we can conclude that  the three   
components of a complete planetary system;
star, dusty debris disc and  planets, increase all together with metallicity and at
the same time the largest massive complete systems are the younger ones. Recently, \cite{ghezzi2018}  also confirmed a similar relation usign  subgiants stars (``retired A stars'').
     
\subsection{Comparisons with debris discs containing larger grains}

As mentioned above, the $M_{\rm d}$ values used in this work were obtained from
mid-infrared (MIR) measurements, and represent discs of dust with radii of some tenths
of AU from the central star and resulting from the emission of grains of
diameters smaller than a millimeter. To detect larger grains of sizes of mm
or more, observations must be made in the submillimeter spectral range. In
this way, more extended and cooler regions of the discs can then be attained. Recent
results of the SONS survey of debris discs in general \citep{holland2017}
by means of measurements of fluxes made mainly at 850 $\mu$m, enable to
obtain disc sizes ranges equivalent to 1 -- 10 times the Kuiper Belt radius
of our Solar System. Also,  small and large grains
could have different evolution. That is, their decline in time due to the
removal of grains from the discs can be different. The SONS survey
\citep{holland2017} observed 49 sources from spectral types from B to M
and measured their dust disc masses.

Now, we explore  the behaviour of the mass of the DD and DDP stars
measured at 850 $\mu$m and we compare them with those measured at the
MIR as presented in this work in the preceding subsections. For this,
we selected the FGK stars observed in the SONS survey. There are
only thirteen objects in common with our work.
Then, for the comparison we use the metallicities values
presented in Table 1. From these thirteen objects, five stars are DDP
and eight are DD stars. All the results and comparisons are presented in the
two panels of  Figure \ref{submili}, where the top panel shows the comparison of $M_{\rm d}$
values at the submm of \cite{holland2017} with the  $M_{\rm d}$  values at the
MIR collected in this work and contained in Table 1. Red triangles are
DDP stars and blue circles represent DD stars. A relative shifted
regular correlation is obtained where systematically $M_{\rm d}$ submm-values
are larger in general than  $M_{\rm d}$ MIR-values. We consider that
this could be expected if extended areas are involved in the submm
discs. In the bottom panel we present the $M_{\rm d}$  submm-values from
\cite{holland2017} versus the metallicity. Even with few points it is
quite notorious that  we found   for measurements at 850 $\mu$m  a smilar result as those measured at
the MIR presented before. We obtain an approximate
similar increasing correlated relation for the DDP objects with metallicity and a
similar uncorrelated distribution for DD objects.
For longer wavelengths as 1.3 mm,  even more extended halos were recently detected around very young DD stars: HD 32297 and HD 61005. These
 detections indicate that mm sized grains are present in these halos \citep{MacGregor2018}.

\section{ Lithium properties in debris disc systems}

\subsection{The lithium depletion}

The element Lithium is depleted in the stellar atmospheres because the original Li, with which the star was formed, is transported by convection to the base of the convection
zone where this element is destroyed. After, the same convection produce an atmospheric Li dilution, by transporting to the surface, internal material poor in Li.
Then both mechanisms contribute to the surface Li depletion \citep{bouvier08}.
The Li depletion mechanism investigated in this work, is based on a strong magnetic disc-star rotation coupling. This coupling induces also  strong internal mixing shears
(\cite{eggenberger2012}, hereafter E12). The  PP disc produces a braking or locking phase that depends only on the lifetime of the disc. By reducing the
external stellar rotation, the efficiency of mixing increases in the transition region between the convective and radiative zones. This mixing becomes the main 
cause for the Li depletion (see Bouvier, 2008 for an initial work on this subject). The E12 model applies for solar-type stars  corresponding to the stars selected 
in this work. One important parameter in the E12 model is what is called ``the end of the disc locking'', which practically determine the disappearance of the effect
of the PP disc. How long does a PP disc live? In general the literature agrees in values up to 10 Myr \citep{williams11} however, larger lifetimes could be possible for some
stars with stellar masses less than 2 $M_{\odot}$  \citep{ribas15,wyatt08}. In general, measured lifetimes of PP discs depend on the Near-IR or Far-IR wavelengths 
considered and on the distances of the radial distribution of the disc mass with respect to the central star.

\begin{table}
\centering
 \caption{Planetary parameters of the DDP sample }
\label{t:planetsDDP}
\small
\begin{tabular}{lcccc}
\hline
\hline
 Planet        &  N$^{\rm o}$  & $M_{\rm planet}$  &   $\rm a$ &     [Fe/H]         \\
 name          &  planets        & ($M_{\rm jup}$)   &    (AU)   &      (dex)            \\
\hline                                                                 
    HD 1461 c       &   2     &    0.017            &  0.1117    &      0.18                 \\
    HD 1461 b       &         &    0.02             &  0.0634    &                      \\
    HD 10647 b      &   1     &    0.93             &  2.03      &      -0.05                       \\
    HD 10700 f      &   2     &    0.012            &  1.3340    &      -0.52                   \\
    HD 10700 e      &         &    0.012            &  0.5380    &                              \\
    HD  20794 b     &  3      &    0.088            &  0.13      &      -0.41                  \\
    HD 20794 d      &         &    0.01             &  0.36      &                             \\
    HD 20794 e      &         &    0.015            &  0.51      &                              \\
    HD 22049 b      &   1     &    3.09             &  3.39      &      -0.15                      \\
    HD 38858 b      &   1     &    0.096            &  1.04      &      -0.22                       \\
    HD 39091 b      &   1     &    1.03             &  3.28      &      0.11                        \\
    HD 40307 c      &   6     &    0.020            &  0.08      &      -0.36                       \\
    HD 40307 d      &         &    0.027            &  0.13      &                         \\
    HD 40307 f      &         &    0.011            &  0.25      &                        \\
    HD 40307 b      &         &    0.012            &  0.05      &                        \\
    HD 40307 e      &         &    0.011            &  0.19      &                        \\
    HD 40307 g      &         &    0.022            &  0.60      &                        \\
    HD 40979 b      &    1    &    4.01             &  0.85      &      0.21              \\
    HD 45184 b      &    1    &    0.04             &  0.06      &      0.04             \\                             
    HD 50499 b      &    1    &    1.71             &  3.86      &      0.22             \\
    HD 50554 b      &    1    &    5.16             &  2.41      &      0.05              \\
    HD 52265 c      &    2    &    0.35             &  0.32      &      0.21             \\
    HD 52265 b      &         &    1.05             &  0.50      &                           \\
    HD 69830 d      &    3    &    0.253            &  0.63      &      0.02                  \\
    HD 69830 c      &         &    0.165            &  0.19      &                           \\
    HD 69830 b      &         &    0.143            &  0.08      &                           \\
    HD 73526 b      &   1     &    2.25             &  0.65      &      0.27                 \\
    HD 108874 c     &    2    &    1.018            &  2.68      &                           \\
    HD 108874 b     &         &    1.36             &  1.05      &               \\
    HD 113337 b     &    1    &    2.83             &  0.92      &      0.2        \\
    HD 115617 b     &    3    &    0.016            &  0.05      &      0.13      \\ 
    HD 115617 c     &         &    0.057            &  0.22      &                \\
    HD 115617 d     &         &    0.072            &  0.48      &                 \\
    HD 117176 b     &     1   &    6.6              &  0.48      &      0.02      \\
    HD 128311 c     &      2  &    4.19             &  1.76      &      -0.06     \\
    HD 128311 b     &         &    2.18             &  1.10      &                 \\
    HD 130322 b     &   1     &    1.05             &  0.09      &      0.01      \\
    HD150706 b      &   1     &    2.71             &  6.70      &      0.01      \\
    HD178911B b     &   1     &    6.292            &  0.32      &      -0.01     \\
    HD187085 b      &   1     &    0.75             &  2.05      &      0.27       \\
    HD192263 b      &   1     &    0.733            &  0.15      &      0.21      \\
    HD210277 b      &   1     &    1.23             &  1.10      &      -0.06      \\
    HD215152 e      &   4     &    0.010            &  0.15      &      0.18      \\
    HD215152 b      &         &    0.006            &  0.06      &                \\
    HD215152 d      &         &    0.008            &  0.0879    &                 \\
    HD215152 c      &         &    0.004            &  0.07      &                \\
    HD216435 b      &    1    &    1.26             &  2.56      &      -0.08     \\
    HD222582 b      &     1   &    7.75             &  1.35      &      0.2        \\
    HD224693 b      &      1  &    0.71             &  0.23      &      0.28      \\
\hline                                                                                 

    \end{tabular} 
{Source: http://exoplanet.eu/catalog/   }
\end{table}

There is however, a point concerning the end of the magnetic braking mechanism by the disc and for which there is not yet a solution. The PP disc lifetime is 
constrained by the duration of the gas in the disc (which is not known precisely), given by the minimum gas in the disc necessary to support a magnetic field 
capable to brake the stellar rotation. Estimation of the duration of the gas component of a PP disc must then be based on other arguments such as: gas accretion 
onto the star, giant planetary formation, stellar chromospheric evaporating winds. Stellar winds can also be capable to brake the star via angular momentum
due to the remotion of momentum \citep{romanova2016,matt10,matt12}.

Winds may originate either in a thin region close to the disc or in the outermost parts of the magnetosphere surrounding the star. In any case, models 
indicate that the power of the wind is given by the accretion. In other words, if there is accretion there will be wind.
In this situation, the braking would be more controlled by
the wind than by the disc itself. Nevertheless, the accretion
exists if there is gas to be accreted. There is then always, a
dependence on the existence of gas for the action of both;
the direct magnetic or the wind braking, mechanisms. Because one of the purposes of this section consists of exploring the rotation-disc connection of the E12 model, we use
here the calculations for the minimum lifetime PP disc of 3
Myr and that of 9 Myr as the maximum considered by this
model. We note that these lifetimes, specially that of 3 Myr, are in agreement with the disc lifetimes of young
stars with intermediate masses \citep{hernandez2005,fairlamb2015}. 
After any maximum PP disc lifetime, this disc pass
by a transition phase \citep{wyatt2015} to then became a
dusty debris disc containing few or practically no gas at least for
FGK stars  \citep{hughes2018,wyatt2018}.

\subsection{The lithium distributions}

\subsubsection{Distribution with mass}

The studied stars have in general stellar masses mainly between 0.7 $M_{\odot}$   and 1.3 $M_{\odot}$.
Stars with $M_{\star}$ $>$ 1.0 $M_{\odot}$  have different levels of a mild or intermediate Li depletion.
This is also the case for some stars of the different groups with $M_{\star}$ $<$ 1.0 $M_{\odot}$  , with the
exception of DDP stars. Nevertheless, strong Li depleted stars are only found among
stars with masses less than one solar mass in agreement with their larger stellar
convection layers. As stellar temperatures mimic well the stellar masses, the Li
distributions with temperatures are similar to those with stellar masses.

 \begin{table}                                                                                                                                                                  
\centering                                                                                                                                                                      
 \caption{Planetary parameters of the CP sample }                                                                                                                               
\label{t:planetsCP}                                                                                                                                                             
\small                                                                                                                                                                          
\begin{tabular}{lcccc}                                                                                                                                                          
\hline                                                                                                                                                                          
\hline                                                                                                                                                                          
 Planet  & N$^{\rm o}$ &   $M_{\rm planet}$   &  $\rm a$ &  [Fe/H]           \\                                                                                                 
 name    & planets     &   ($M_{\rm jup}$)     & (AU)   & (dex)            \\                                                                                                   
\hline                                                                                                                                                                          
1237 b    & 1  &  3.37	  &  0.49	&  0.09                  \\                                                                                                            
3651 b    & 2  &  0.231	  &  0.29	&  0.12                  \\                                                                                                            
3651 c    &    &  0.18	  &  0.04       & 	                    \\                                                                                         
4308 b    & 1  &  0.040   &  0.11	&  -0.34                   \\                                                                                             
                                                                                                
10697 b   & 1  &  6.83	  &  2.16	& 0.18                    \\                                                                                                       
13445 b   & 1  &  4.01	  &  0.11	& -0.24                    \\                                                                                                      
17051 b   & 1  &  2.26	  &  0.92	& 0.12                    \\                                                                                                       
23079 b   & 1  &  2.45	  &  1.59	& -0.08                    \\                                                                                                      
28185 b   & 1  &  5.7	  &  1.03	& 0.23                    \\                                                                                                     
33564 b   & 1  &  9.1	  &  1.1        & 0.12                   \\                                                                                                              
72659 b   & 1  &  3.15	  &  4.74	& -0.11                   \\                                                                                                  
75732 f   & 5  &  0.147   &  0.77       & 0.33                    \\                                                                                                        
75732 e   &    &  0.025   &  0.01       &                    \\                                                                                                                      
75732 d   &    &   3.86   &  5.44	&                          \\                                                                                                               
75732 c   &    &  0.178   &  0.23	&                           \\                                                                                                              
75732 b   &    &  0.84    &  0.11	&                           \\                                                                                                    
95128 d   & 3  &  1.64	  &  11.6	&  0.06                   \\                                                                                                   
95128 c   &    &  0.54    &  3.6 	&                          \\                                                                    
95128 b   &    &  2.53    &  2.1	&                            \\                                                                  
102365 b  & 1  &  0.05	  &  0.46	&  -0.35                    \\                                                                   
114729 b  & 1  &  0.84	  &  2.08	&  -0.33                   \\                                                                    
115383 b  & 1  &  4.00	  &  43.5	&  0.24                    \\                                                                    
134987 b  & 2  &  1.59	  &  0.81	&  0.31	                   \\                                                                    
134987 c  &    &   0.82   &  5.8        &                           \\                                                                   
136352 b  & 3  &  0.016   &  0.09	&  -0.28                   \\                                                                    
136352 c  &    &  0.035   &  0.16     &                               \\                                                                
136352 d  &    &  0.03    &  0.41	&                       \\                                                                       
145675 b  & 1  &  4.64	  &  2.77	&  0.43                    \\                                                                    
147513 b  & 1  &  1.21	  &  1.32	&  0.11	                  \\                                                                     
154088 b  & 1  &  0.019   &  0.13        &  0.28	                   \\                                                                    
154345 b  & 1  &  1.3	  &  4.3	&  -0.1	                  \\                                                                     
160691 b  & 4  &  1.676	  &  1.5	&  0.33                    \\
160691 c  &    &  0.03    &  0.09       &  		                    \\
160691 d  &    &  0.52	  &  0.92       &  		                       \\
160691 e  &    &  1.814	  &  5.23       &  		                    \\
189567 b  & 1  &  0.031   &  0.10	&  0.24	                    \\
189733 b  & 1  &  1.142	  &  0.03	&  -0.02                     \\
192310 c  & 2  &  0.076	  &  1.18	&  -0.03                    \\
192310 b  &    &  0.053   &  0.32        &   	                    \\
195019 b  & 1  &  3.7	  &  0.13	&  -0.01                     \\
196050 b  & 1  &  2.83	  &  2.47	&  0.34	                    \\
196885 b  & 1  &  2.98	  &  2.6	&  0.15                    \\
213240 b  & 1  &  4.5	  &  2.03	&  0.16	                    \\
216437 b  & 1  &  1.82	  &  2.32	&  0.24	                   \\
217014 b  & 1  &  0.47	  &  0.052	&  0.15	                    \\
\hline   
    \end{tabular}  

{Source: http://exoplanet.eu/catalog/   }\\

\end{table}

    \begin{figure}
     \centering
  \includegraphics[width=7.5cm]{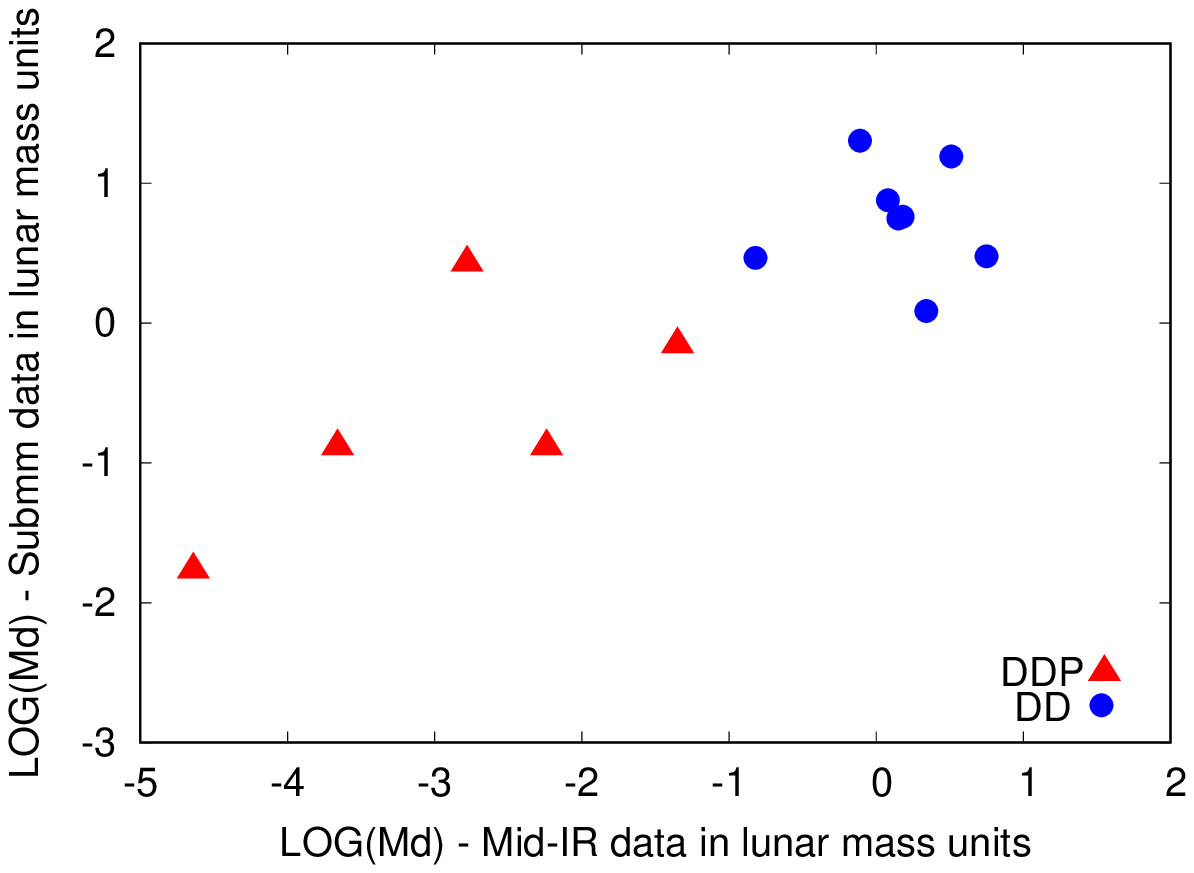}\par     
  \includegraphics[width=7.5cm]{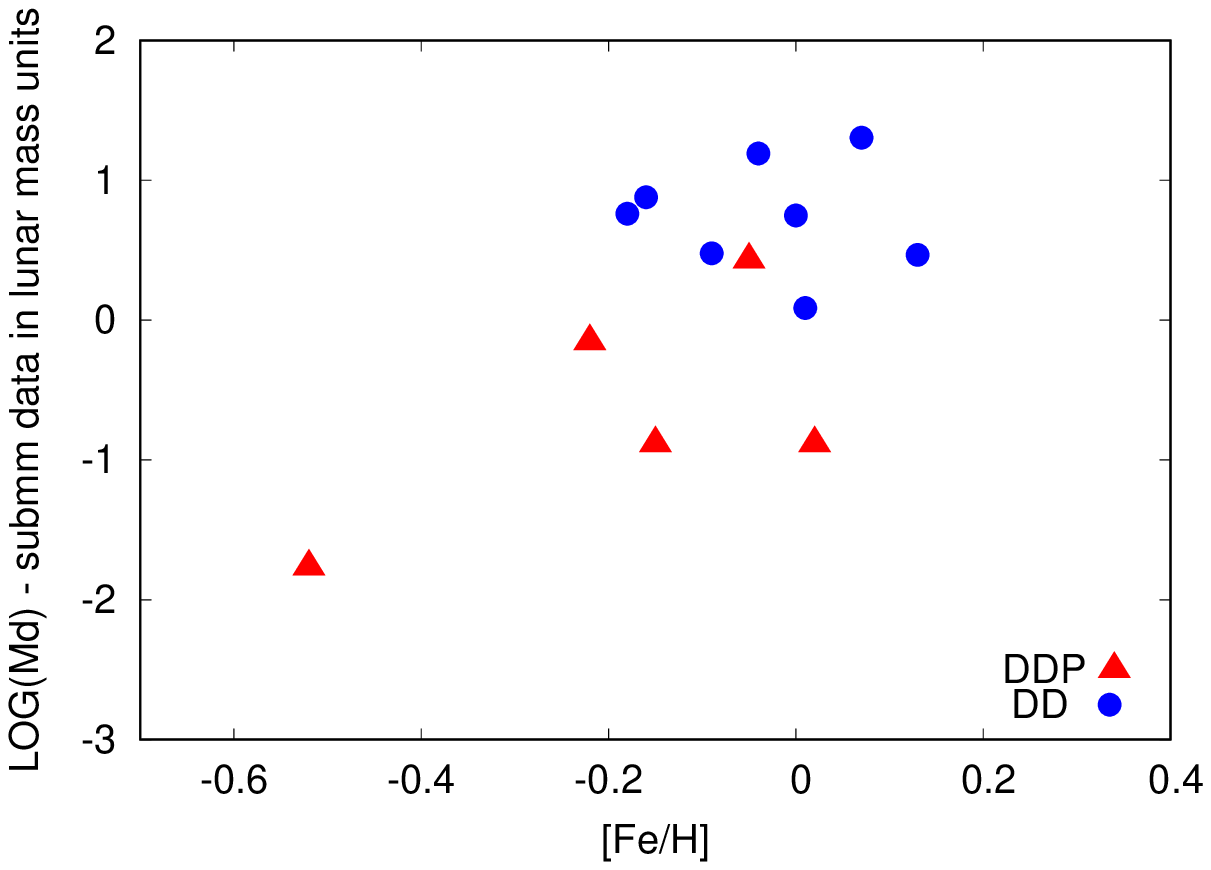}\par     
   \caption{ Top Panel: comparison of $M_{\rm d}$
values at the submm taken from Holland et al. 2017 with the  $M_{\rm d}$  values at the
MIR collected in this work (see Table \ref{t:parameters0}). The DDP common stars are: HD10647
HD22049, HD38858, HD115617 and HD10700. The DD common stars are: HD377
HD61005, HD104860, HD107146, HD170773, HD181327, HD191089 and HD207129. Bottom Panel: $M_{\rm d}$  submm-values from
Holland et al. 2017  versus the stellar metallicity. }
    \label{submili}
    \end{figure}

 \subsubsection{Distribution with rotation}

The stars belonging to the four groups are in general slow rotators with $\mathfrak{v}\sin i$ $<$ 5 km s$^{-1}$.
Some few relative younger stars have faster rotation velocities only in the C and DD groups. As expected, these fast rotating stars
have not been braked sufficiently and their Li abundances have been maintained relatively high as is the case of stars HD 693, HD 133295, HD 181321 and HD 35296.

 \begin{figure*}
\centering
   \includegraphics[width=15cm]{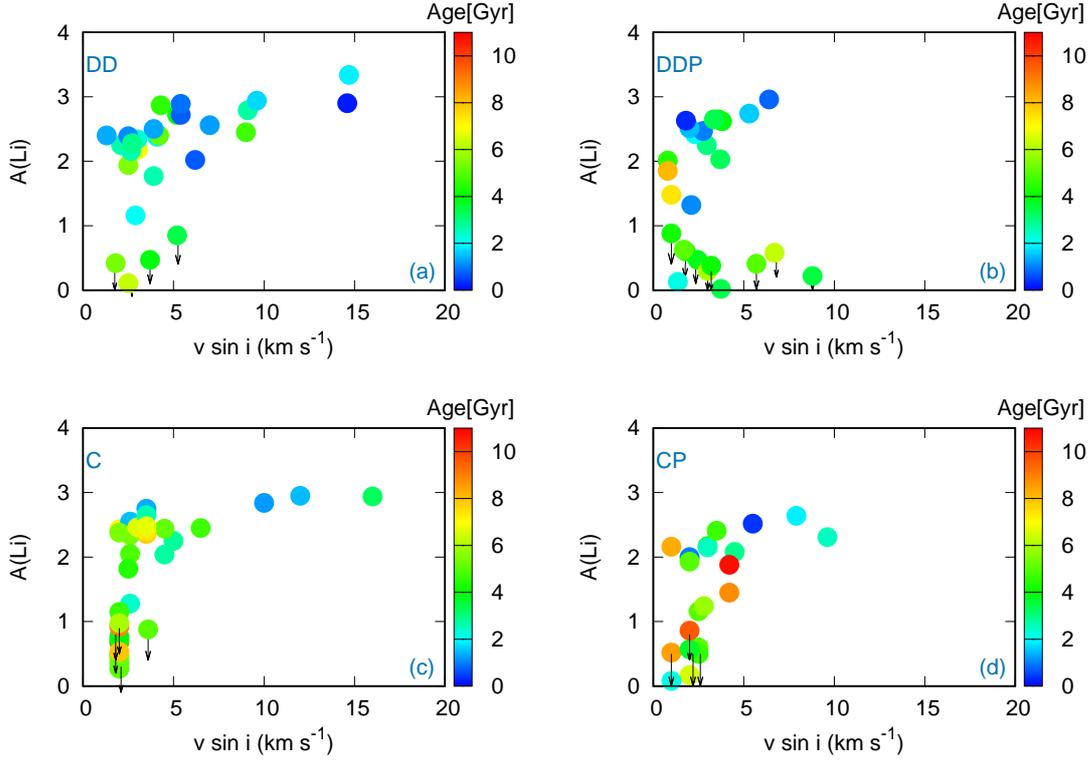}
    \caption{ Lithium abundances vs  $\mathfrak{v}\sin i$, for the four groups, the color bar indicates the age parameter in Gyr.   Note that the most of the values of A(Li)$<$1
    correspond to  upper limit values (see Table \ref{t:parameters0}). In addition,   for slow rotating stars,  it is only possible to
obtain upper limit values for this parameter, this is $\mathfrak{v}\sin i$ near to 2.5 --3.0  km s$^{-1}$ for FEROS spectrograph.   
}
   \label{li-rotationli}
    \end{figure*}

\begin{figure*}
    \centering
    
\includegraphics[width=0.7\columnwidth]{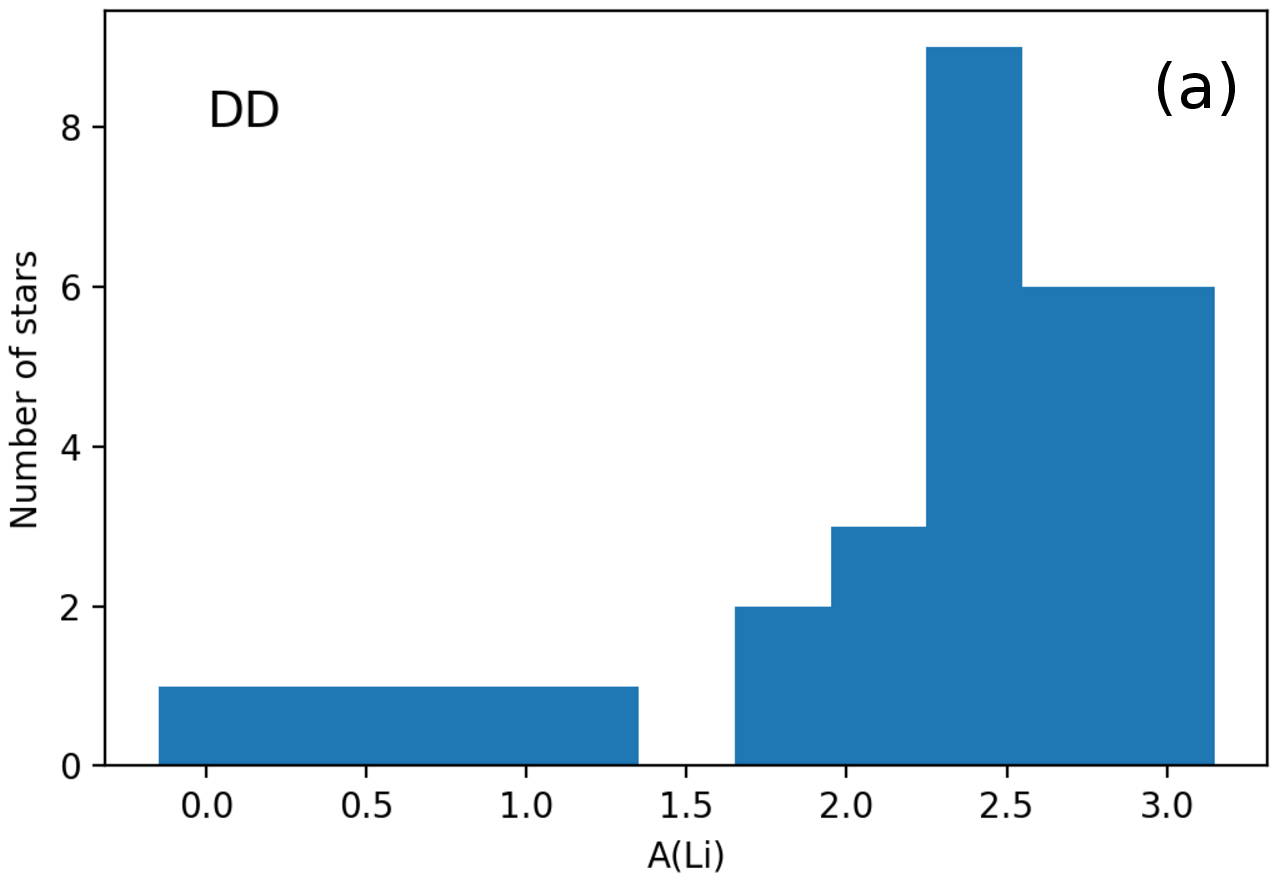}
\includegraphics[width=0.7\columnwidth]{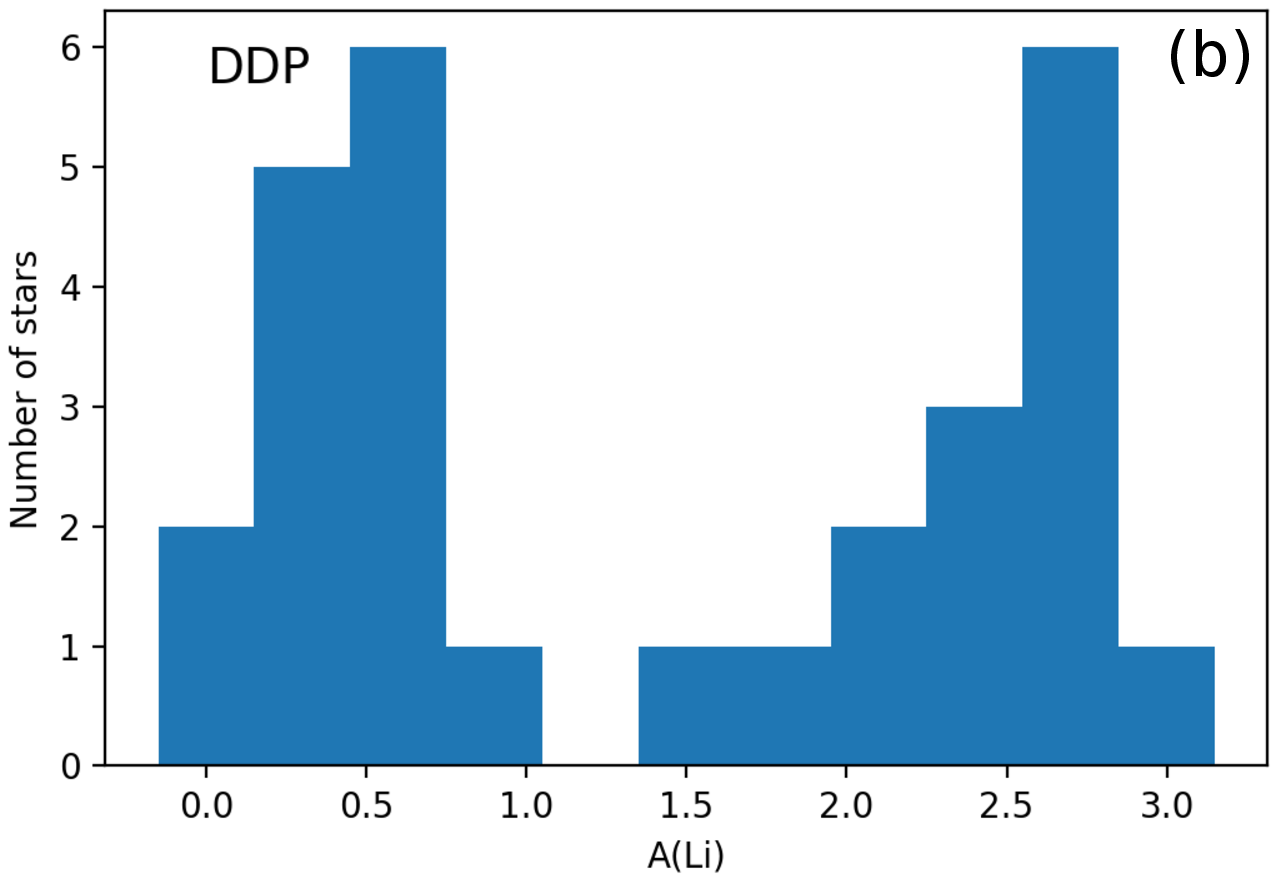}
\includegraphics[width=0.7\columnwidth]{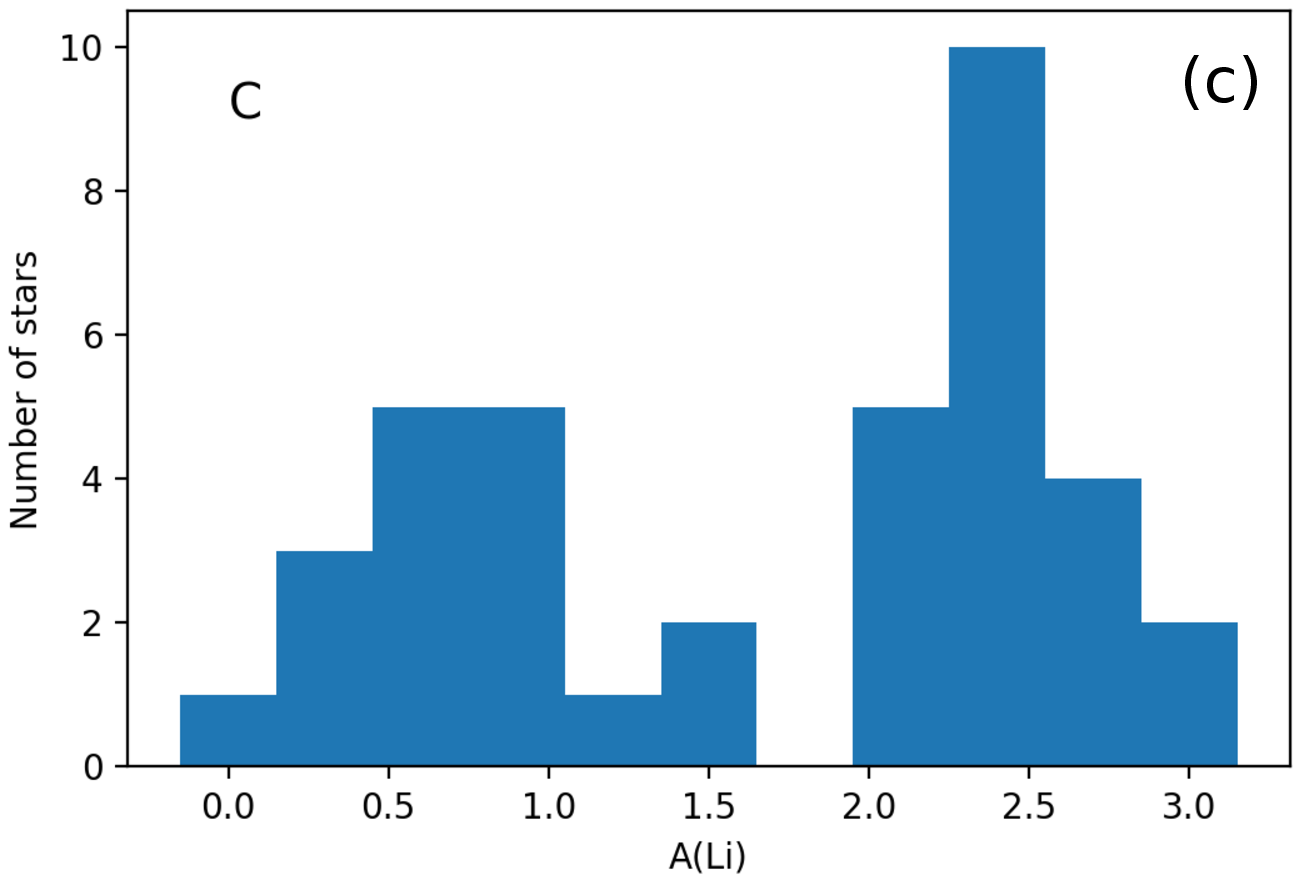}
\includegraphics[width=0.7\columnwidth]{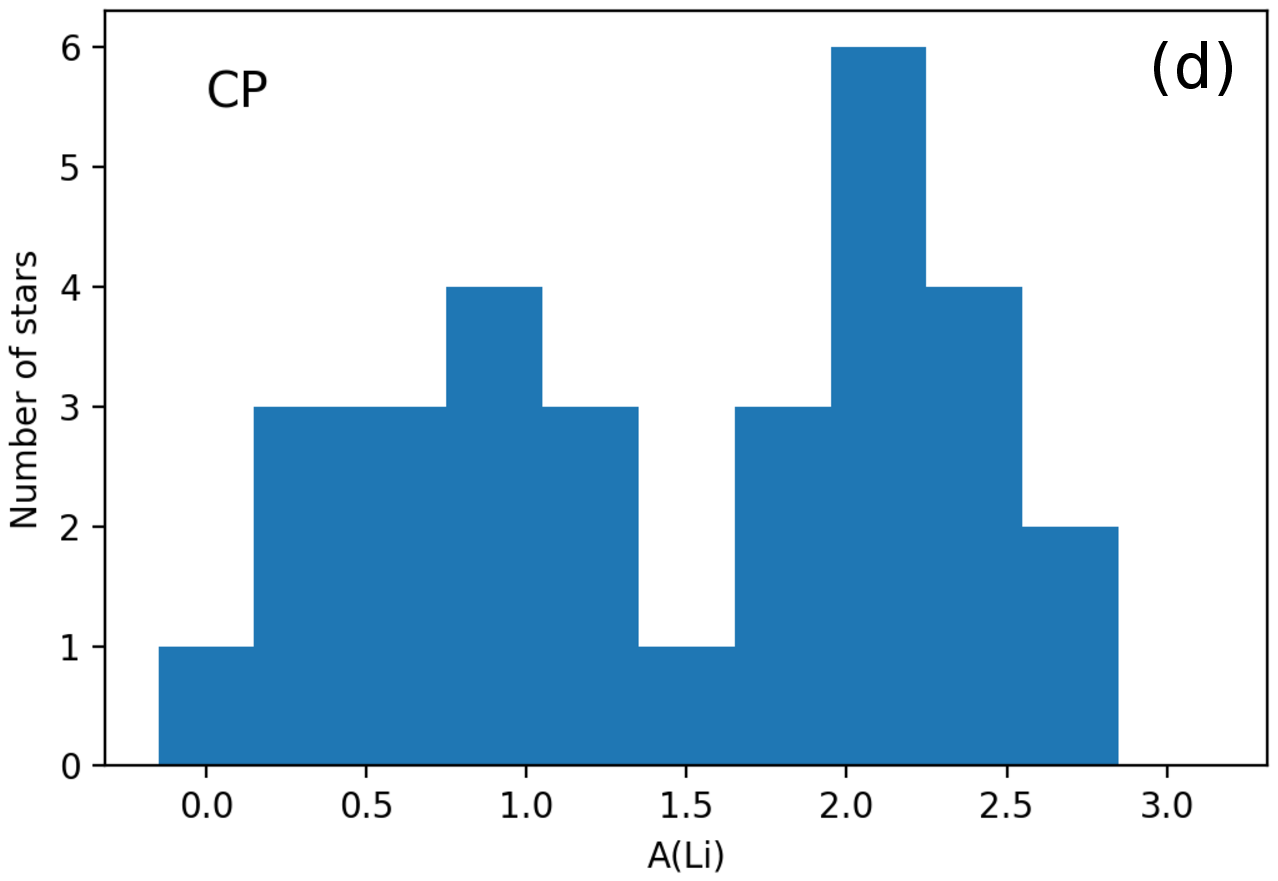}
\caption{  
Distribution of lithium abundance for the four groups.
These histograms follow the history of the Li depletion in the four groups 
of stars considered in this work. One part of the stars of each group remained with the Li 
abundances resulting from the Li depletion during the PP phase up to terminal values of A(Li)  $\sim$ 2.2. 
Other stars continued with a very slowly Li depletion during the main-sequence, up to terminal
 values of A(Li) $<$ 1.0.   For all panels a c and d representing  DD, C
and CP groups respectively, there exist a maximum PP depletion peak between A(Li) $=$ 2.2 and
2.4 which correspond to the action of a PP disc with a lifetime of 9 Myr satisfying the 
prediction of model E12. The depletion behaviour for the DDP group is different and 
have not an easy interpretation with the E12 model (see text). }
\label{histo}
    \end{figure*}

 All panels of Figure \ref{li-rotationli} show that there is a concentration of the observed Li abundances around the predicted terminal depleted Li
 values of  model E12 of A(Li) $\sim$ 2.2. These concentrations  are better represented in the panels of Figure \ref{histo}. Model E12 predicts that
 after the action of the longest  disc lifetime of 9 Myr (see fig 8 in E12) the terminal Li depleted abundances are equal to A(Li) $=$ 2.3.
 As can be  seen in the panels a, c and d of Figure \ref{histo}   there are three clear  concentrations of the depleted values of A(Li) at the value of $\sim$ 2.2 
 for the group CP and at $\sim$ 2.4 for  DD and C groups. Those two values   are equidistant to the value corresponding  to the mentioned action of 
 disc with a lifetime of 9 Myr.  These three final concentrations  agree with the E12 model. On the contrary,   the behaviour of the Li depletion is 
 different for  the DDP group.  Figure \ref{histo}-b shows that there is no final accumulation peak of depleted values   of A(Li), but an 
 oppositely behavior showing the absence of a final accumulation peak.

 In any case, by considering all the measured A(Li) values larger than 2.0, we are confirming the E12 model predictions which are valid for stars with
masses near the  solar mass.  However, it is more difficult to explain the  values of  A(Li) $<$ 2.0  for later ages.

It is expected that a slow Li depleting mechanism would be acting during the whole MS stage, due to an internal mixture mechanism maybe depending to a certain degree,
to that generated during the PP disc phase.
In this case, the difficulties depend not only on the stellar internal response to the short spin up rotation surface process between $\sim$ 10 to $\sim$ 40 Myr 
\citep{bouvier08}, but also to the long spin down surface phase provoked by magnetised winds \citep{johnstone2015}. To our knowledge, there is not yet a quantitative
theory that predicts this long term evolution, which also depends on the transport mechanism in the radiative zone.  Maybe an appropriate theory, not existing
yet, could explain the observed distribution of the very low Li depleted values shown at the left side of each panel of Figure \ref{histo}.

In any case, we observe  that the majority of highly depleted Li stars have masses less or equal to one solar mass (see Table 1) with larger convective zones and
with ages between 3 Gyr and $\simeq$ 10 Gyr. Even with a large dispersion, the Li abundances decrease with increasing age. This known property in the literature can be 
seen in all panels of Figure \ref{li-rotationli}.

\subsection{Lithium in debris discs stars with and without planets}

\begin{figure*}
\centering
   \includegraphics[width=15cm]{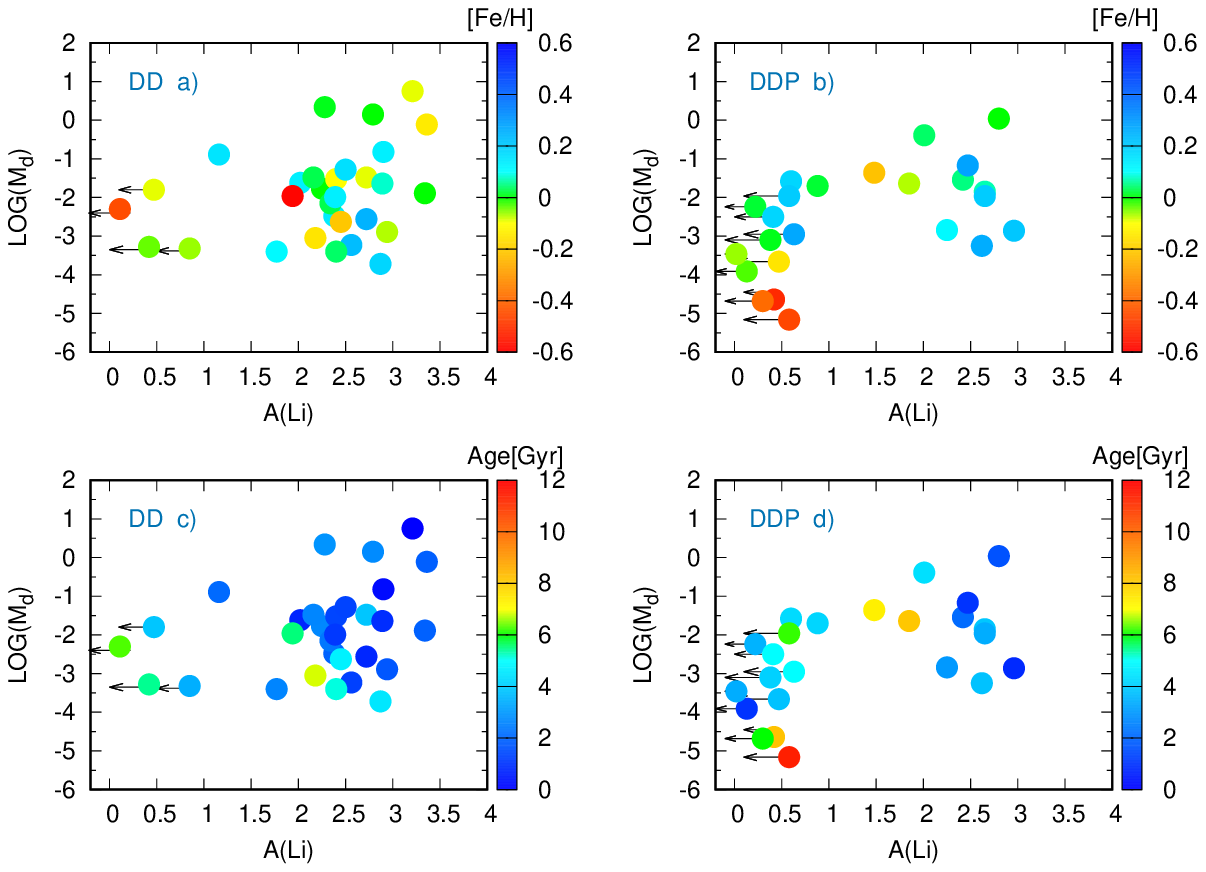}
    \caption{ Lithium abundances vs LOG $M_{\rm d}$ for DD and DDP groups, in panels a) and b), the color bar indicates the stellar metallicity, whereas in panel c)
    and d) the color bar indicates     the stellar ages in Gyr. The values corresponding to A(L) $<$ 1.0 are uper limit values. Error bars of the rest of the sample are the size of the symbols except in one point as is in detail in column 11 of Table \ref{t:parameters0}.  }
   \label{li-mdx4}
    \end{figure*}

In Section 3 we have explored the ensemble of metallicity properties of DD and DDP stars. We can ask now what is the
behaviour of the stellar Li abundance in these same objects? In
Figure \ref{li-mdx4}  is presented the distributions of the dusty disc masses
$M_{\rm d}$ in function of the Li abundance for DD and DDP stars
considering also colour scales for metallicity and age. 

In both panels (a) and (c) of Figure  \ref{li-mdx4} representing DD stars, we can see that
a very large dispersion of $M_{\rm d}$ values is present with any correlation
with the Li values. However, a relative coherence appears in panel
c in respect to ages. In reality, this shows the fact that a large part
of DD stars are younger than the other groups (see Section 3.1).
Whatever, when metallicity is considered as in panel (a), we can see
a complete disorderly distribution. On the contrary, DDP stars
distributions of $M_{\rm d}$ versus Li abundances (panels (b) and (d)) appear,
in the both cases of metallicity and age, as presenting a certain
order. It is specially notorious the large quantity of highly Li
depleted objects (A(Li) $<$ 1.0) and the relative orderly distributions
of intermediate Li depleted stars (A(Li) $>$ 1.0). In reality, all these
$M_{\rm d}$ - Li distributions are a direct reflection of the distributions of $M_{\rm d}$ values in
function of the stellar mass as seen in both panels of Figure \ref{DD}. In fact,
all these distributions of Figures \ref{DD} and \ref{li-mdx4} are
similar, showing that the Li abundances follow directly the stellar
masses (and indirectly the size of the convective stellar zone ) as
must be the case. In conclusion, very different distributions of $M_{\rm d}$
values in function of the Li abundances of DDP stars and of DD
stars exist. They appear to be the consequence respectively of the
fulfil or not of the two conditions to form gaseous planets as
discussed in Subsection 3.2.3.

\subsection{Is the presence of giant planets an indication of larger lithium depletion?}

The question, if the presence of extrasolar giant planets might produce a larger Li depletion relative to stars with apparently
no detected planets, have resulted, since the year 2004 in a quite large number of works presenting opposite results. Those advocating
for larger Li depletion in stars with giant planets are: \cite{israelian04, takedaKawanomoto2005, chenZhao2006,israelian2009, takeda2010,sousa2010,gonzalez2010,gonzalez2014,delgadomena2014,figueira2014}. On the contrary, those who found no connection between
the presence of planets and the Li  abundance, are: \cite{ryan2000,luckHeiter2006,baumann2010,ghezzi2010b,ramirez2012,bensby2018}.
If we exam panels a and b, or d and c of Figure \ref{li-mdx4} representing the mass of dusty discs versus the stellar Li abundances of DD and DDP stars,
we can see that the distribution of the strongly Li depleted (A(Li) $<$ 1.0 ) stars are very different.

A much larger number of strongly Li deficient objects are present among the DDP stars, than in DD stars. In a first impression this could mean that
the presence of planets would produce a stronger Li depletion on DDP stars in respect to DD stars. However, this can only be an age effect and not
an effect due to the presence of planets. In fact, the mean age of the stronger Li depleted DDP is 4.6 Gyr and this age is almost the same mean age
of 4.1 Gyr of the same strongly Li depleted DD stars. In summary, even if the intermediate Li abundances (A(Li) $>$ 1.0 ) DD stars, with a mean age
of 2.4 Gyr are younger than the corresponding DDP stars, with a mean of 3.2 Gyr, both groups of stars need to attain the same age near 4.6 Gyr to be
completely Li depleted by a normal slow main-sequence depletion mechanism. Then, the presence of planets appears not to be the main cause of Li depletion.

The above mentioned discussions in the literature refers however, to a comparison between CP and C stars. In the absence of $M_{\rm d}$ values for these stars,
we can nevertheless, compare directly the mean ages for these stars for which they attain A(Li) values less than 1.0. These  mean ages are respectively for
CP and C stars equal to 6.1 Gyr and 5.8 Gyr. Both ages being relatively similar, we can conclude that also here, an age effect could be the determinant
effect to deplete largely the Li abundance and not the presence of planets.

\section{Results and Conclusions}

 One of the main purposes of this work consists to  deepen the  understanding of the metallicity properties of dusty debris discs stars, 
 with and without planets. Moreover, we also studied,
 for the first time the  lithium distributions and properties of these stars.
  For this study,  we have selected four different groups of field solar FGK type stars, avoiding any
 close binary stars and considering only main sequence stars. The selected groups of stars have
 the following properties: the first, containing only debris discs  with no detected
 planets. The second, contains DD and  planets (DDP). The third, containing
 only detected giant planets (CP) and the fourth, stars without detected discs or planets
 (C). A large part of the objects have been observed and coherently measured by us.

 One important tool employed in this study consists to use the dust masses of the discs
 ($M_{\rm d}$). We assume  that the metallicity of the dusty discs is the same as that of
 their central stars. Concerning the DD stars, it was already known by several authors,
 using indicators other than $M_{\rm d}$, that stellar metallicity and the presence of debris discs does not correlate.
 Recently, \cite{gaspar2016} using $M_{\rm d}$ values calculated by them,
 explored the  correlation between  dust masses  debris discs with the metallicities of their central stars for a very large number of objects .
 
 Nevertheless, their use of a mixed list of objects containing upper limits
 IR luminosities, binaries and no distinction of those objects having planets or not, impeded to
 obtain a clear result. In this work, using a clean list of  DD stars, we confirm the
 completely independence of $M_{\rm d}$ values in respect to the metallicity.

 In fact, we also confirm that the stellar masses of the DD sample, are also
uncorrelated with the metallicity. We then conclude that the whole masses (stellar and
disc) of DD stars are uncorrelated to metallicity. On the contrary the masses of stars
containing giant planets, as DDP and CP objects, are correlated with metallicity. We
present  a new result by showing that DDP objects are those presenting the
stronger correlation with [Fe/H]. We also found that stars containing giant planets as
DDP and CP stars have a different distribution of their stellar masses compared to
stars do not showing giant planets. Whereas DDP and CP stars have a flat distribution
of stellar masses between $\sim$ 0.7 and 1.3 M$_{\odot}$, DD and C stars present a peak at very
approximately one solar mass. This result indicates why in general, stars need to have
higher masses, then higher metallicities, to form giant planets.

 One aspect  discussed in this work concerning DD stars, is their asymmetrical distribution in
respect to metallicity. We confirm  an important deficit of metallic deficient DD
stars. This deficit was already noted by \cite{maldonado12}, \cite{montesinos2016} and \cite{gaspar2016}.
On the contrary, the population on the metal rich side of the distribution is
quite full. We propose  a scenario that could explain this distribution of DD
objects. These DD stars, which have a mean age  of 2.5 Gyr are very
much younger than C stars with a mean age of 4.8 Gyr. We can  then suggest that  DD stars could be
transformed into C stars when they loose their dusty discs. In fact, the very few 
metal deficient DD stars detected  are very old and considered to be the last
survivors.

Concerning DDP stars, we found that their dust discs masses are directly related to the
corresponding stellar masses. This relation is similar to the same relation found
in very young stars systems with ages less or equal to $\sim$ 10 Myr. Independently of
this, we also found that the three components of the DDP systems; stars,   dusty discs and cores of
planets, have each their masses directly dependent on metallicity. That is a relative
continuous increase of their masses with an also continuous increase of the
metallicity scale.
The new result obtained, in which the disc  masses of DDP stars appear to be correlated to metallicity is one of the most important results of this work.

The increase of the planetary masses with the metallicity  shown in this work was already known in the literature.
Collecting all these properties among the DDP stars, we can infer that for
each increasing value of the metallicity, the complete stellar system (star, dust and
planets) have successively larger total masses. 

Also, we note  that these successively
larger massive systems, are also successively younger. To understand these
properties, we consider that two natural conditions are necessary to be fulfilled for a
system to form giant gaseous planets; one, to dispose of a sufficient metallicity and
other, to dispose of a sufficient total (gas and dust) massive PP disc. DD stars do not
form giant gas planets because, even if they can dispose of a sufficient metallicity, they do
not fulfil the second condition of having a PP disc total mass, larger than a critical
minimum mass to form a giant gaseous planet.

One aspect that has not been explained in this work concerning
the dusty discs masses, is why the $M_{\rm d}$  values of DD stars are
somewhat larger in general than those of DDP stars (see Figures 4
and 10). 
This difference is even present in Figure \ref{submili}
representing the submillimetric measurements of
extended discs with larger sized particles. We don't know if we
have a possible answer for this, but it appears that it is not
related to the size of the involved grains. In fact, in Section 3.3
it is found that DDP and DD stars present the same properties
and behaviours, of order and disorder, respectively in the
submm regime, as is the case for the smaller grains in the MIR.

We can however, explore others ways to eventually find a
different intrinsic nature between DD and DDP stars, in order
to try to explain their dusty disc mass differences. We can
consider for instance, that the PP discs of DD stars could have an
original minor gas-to-dust ratios than those corresponding to DDP
stars. In this scenario, dust aggregative collisions will be
dominant \citep[][and references therein]{testi2014}  and could
maybe form an excess of planetesimals. This excess could
then produce, by collisions, more dust in the subsequent
debris disc stage forming this way, the more massive dusty
discs found for DD stars. However, this scenario must contain
a mechanism that somehow avoid the formation of large
planetary cores in order to explain the absence of giant
gaseous planets in DD stars.  An alternative scenario could be that eccentric giant planets are efficient in
cleaning the disc \citep[see e.g.][]{raymond2011,maldonado12}.

Similarly, as the proposed evolution in which DD stars turn to be C stars
at the end of their evolution during the MS, when their dusty discs have largely
vanished (at least part of C stars can have this origin), we can also suggest that DDP stars are
transformed into CP stars by the same process of loosing their dusty discs. In fact,
the determined mean ages for these objects are appropriate for this kind of evolution.
The respective mean ages for DD and C stars are 2.5 Gyr and 4.8 Gyr whereas those
of DDP and CP stars are respectively 4.1 Gyr and 5.4 Gyr.  We must note however, that the larger age of stars with
planets could be somewhat biased towards older stars. This because
the first exoplanet surveys tried to avoid very active young stars. 
 
 Concerning the Li depletion, we adopted the stellar model \citep{eggenberger2012} based on a strong magnetic coupling between the star 
 and the PP disc. We found that the observed Li abundances for C, DD and CP stars are in agreement with the depletion predicted by the model with a 9 Myr lifetime.
 The Li depletions for the expected more massive PP discs of DDP stars
 behaves differently and we do not find a clear interpretation for them using this model.

The study of the distribution of the Li abundances in DDP and DD stars is very
instructive. Both distributions are very different. On DDP stars the intermediate Li
depleted stars with A(Li) $>$ 1.0 are relatively orderly distributed. Their Li abundances
correlate with an increased metallicity and a decreased age.

A direct comparison of the distributions of the dusty debris discs
masses $M_{\rm d}$ of DD and DDP stars, in function of the Li abundances
and the same in function of the stellar masses, show that they are
similar. This indicates that the Li abundances follows well the stellar
masses.

Finally, to try to answer the long debated question
if the presence of giant planets produces an increase of the Li depletion we found that
apparently the age, and not the presence of giant planets, is the main cause of the Li
depletion in both cases; DDP stars in respect to DD stars and also in CP stars in
respect to C stars.

\section*{Acknowledgements}
C.CH. and C.G. acknowledges support from  SECYT/UNC and CONICET, L.G. would like to thank the financial support from the Coordena\c{c}\~{a}o de
Aperfei\c{c}oamento de Pessoal de Nivel Superior (CAPES), F. L de A. thanks the support from the Faculty of the European Space Astronomy Centre (ESAC).
R. de la R. thanks Dr. P. Eggenberger for dialogues on the initial preparation of this work and to Syna Snoek for inspiring conversations.
This research has made use of NASA$'$s Astrophysics Data System. We thank the anonymous referees whose comments and suggestions have improved this manuscript.

\bibliographystyle{mnras}
\bibliography{chavero2019} 


\bsp	
\label{lastpage}
\end{document}